\documentstyle[aps,version2,pre]{revtex}    
\begin{document}

\draft

\begin{title}
Complete Pseudohole and Heavy-Pseudoparticle Operator Representation\\ 
for the Hubbard Chain                       
\end{title}

\author{J. M. P. Carmelo and N. M. R. Peres} 
\begin{instit}
Department of Physics, University 
of \'Evora, Apartado 94, P - 7001 \'Evora Codex, Portugal
\end{instit}
\receipt{20 December 1996}
\begin{abstract}
We introduce the pseudohole and heavy-pseudoparticle operator 
algebra that generates {\it all} Hubbard-chain eigenstates from 
a single reference vacuum. In addition to the pseudoholes
already introduced for the description of the low-energy
physics, this involves the heavy pseudoparticles 
associated with Hamiltonian eigenstates whose energy spectrum 
has a gap relatively to the many-electron ground state. 
We introduce a generalized pseudoparticle perturbation theory
which describes the relevant finite-energy ground state transitions.
In the present basis these excitations refer to a small density of 
excited pseudoparticles. Our operator basis goes beyond the Bethe-ansatz
solution and it is the suitable and correct starting point for the study 
of the finite-frequency properties, which are of great relevance
for the understanding of the unusual spectral properties detected in 
low-dimensional novel materials.
\end{abstract}
\renewcommand{\baselinestretch}{1.65546}   
\renewcommand{\baselinestretch}{1.656}   


\section{INTRODUCTION}

The unusual spectral properties of one-dimensional electronic 
quantum liquids \cite{Solyom,Meden} imply that they cannot be 
described by one-electron, Fermi-liquid-like models 
\cite{Pines,Baym}. Further, descriptions of these quantum systems 
in terms of exotic excitations such as holons and spinons 
\cite{Anderson,Essler}, or pseudoparticles 
\cite{Carmelo92,Carmelo93,Carmelo94}, have been limited to the 
low-energy Hilbert subspace, and both bosonization \cite{Solyom,Meden}
and conformal-field theory techniques \cite{Belavin,Blote,Affleck,Frahm} 
also apply only in that limit. However, low-energy studies cannot 
describe the finite-frequency properties, which
are of great interest for the understanding of the unusual
properties detected in real low-dimensional novel materials.
Thus developing a microscopic operator description of these systems 
for arbitrary energies remains an important theoretical challenge. 

For more than sixty years the ``Bethe ansatz'' (BA) \cite{Bethe} 
has played a central role in the analytic solution of a variety of 
``integrable'' many-body problems in condensed matter physics and 
quantum field theory \cite{Bethe,Yang,Thacker,Korepinrev,Lieb}. 
Techniques based on the exact BA solutions were used to describe 
excited states, thermodynamics, and low-energy correlation functions 
\cite{Frahm,Thacker,Korepinrev,Takahashi} and remain an active 
subject of study today. However, the BA solution only
provides limited information on the many-particle problem.
For instance, it does neither provide off-diagonal operator
expressions nor matrix elements. 

In this paper we introduce a suitable pseudohole and 
heavy-pseudoparticle operator algebra which generates {\it all} 
Hubbard-chain \cite{Lieb,Takahashi} eigenstates from a single 
pseudohole and heavy-pseudoparticle vacuum. 
This is the pseudohole vacuum already introduced in Ref. 
\cite{Carmelo96} for the low-energy Hilbert subspace spanned
by the lowest-weight states (LWS's) and highest-weight states 
(HWS's) of the $SU(2)$ $\eta $-spin and spin algebras 
\cite{Heilmann,Yang89,Korepin,Ostlund}
of type I. (See below the definition of types I and II
LWS's and HWS's.) That vacuum is the half-filling and 
zero-magnetic-field ground state (GS). 

Recent investigations have established that although not themselves 
complete, the Hamiltonian eigenstates which refer to the BA
solution can be extended to form a complete 
set of states for the one-dimensional Hubbard model \cite{Korepin}. 
However, the vital and to date open issue of how to describe 
specifically those excitations that dominate response and 
transport at finite energies in integrable quantum liquids and 
related models remains an open problem. Our generalized operator 
algebra is complete and allows the construction of a
pseudoparticle perturbation theory which is the correct starting 
point to solve this problem. The main point is that the
GS transitions which determine and dominate response 
and transport at finite energies involve a small density of
excited pseudoparticles only. Therefore, the description of
these relevant transitions is within the range of our perturbation 
theory. In this paper we introduce the generalized operator
representation. The application of our approach to study the 
finite-energy spectral and transport properties of the Hubbard 
chain will be presented elsewhere \cite{Carmelo96b,Carmelo96d}. 

In contrast to three-dimensional Fermi liquids \cite{Pines,Baym}, 
the low-energy excitations of one-dimensional many-electron 
quantum problems are at zero magnetic field characterized by 
charge - spin separation \cite{Solyom,Meden,Anderson,Essler}. 
This can be interpretated in terms of holon and spinon modes 
\cite{Anderson,Essler,Carmelo96}.
The BA solution of the Hubbard chain \cite{Lieb} at zero magnetic 
field and chemical potential has allowed the identification 
and study of the holon and spinon excitations and corresponding 
symmetry transformations \cite{Essler,Carmelo96}. (The spinon 
excitations of the Hubbard chain are similar to the corresponding 
spinon excitations of the spin $1/2$-isotropic Heisenberg chain 
\cite{Faddeev}.) 

On the other hand, at finite magnetic fields the charge and
spin separation is replaced by a more exotic $c$ and
$s$ separation \cite{Carmelo94}. Here $c$ and $s$ refer
to orthogonal small-momentum and low-energy modes which
couple both to the charge and spin channels. These
can be described in terms of pseudoparticle and
pseudohole excitations \cite{Carmelo96}. In Ref. 
\cite{Carmelo96} it was shown that the zero-magnetic
field holon and spinon excitations are limiting cases
of the general pseudohole excitations.
This has allowed to obtain the exact expression of the 
electron operator in terms of holon and spinon operators 
\cite{Carmelo96}. This study has also provided the relation 
between Hamiltonian symmetry and the transformation of the 
elementary excitations at finite magnetic fields.  
(At finite magnetic fields the usual holon and spinon picture breaks down
\cite{Carmelo96}.) 

The present pseudohole and heavy-pseudoparticle operator algebra is a 
generalization of the low-energy pseudoparticle operator 
representation of Refs. \cite{Carmelo92,Carmelo93,Carmelo94} 
and pseudohole representation of Ref. \cite{Carmelo96}.
For the general case of the Hubbard chain at finite magnetic field 
and chemical potential both the non-LWS's and non-HWS's of the 
$\eta$-spin and spin algebras have energy gaps relative to the 
corresponding canonical-ensemble GS
\cite{Carmelo96,Carmelo95}. (The expressions of these gaps
are determined in this paper.)
The LWS's and HWS's of these algebras
can be classified into two types, the states I (or LWS's I and
HWS's I) and the states II (or LWS's II and HWS's II). While
the Hamiltonian eigenstates I are described only by real
BA rapidities, all or some of the rapidities associated
with the eigenstates II are complex and non real. Since
for finite magnetic field and chemical potential the
states II have energy gaps relative to the corresponding
canonical-ensemble GS, in that case the low-energy physics
is exclusively determined by the states I \cite{Carmelo96}.
Our algebra refers to the whole parameter space
and is expressed most naturally in terms of operators describing 
the pseudohole excitations introduced in Ref. \cite{Carmelo96}, the new 
heavy-pseudoparticle excitations, and the associate 
topological momentum-shift excitations.

We demonstrate that not only the familiar BA solution of the
Hubbard chain \cite{Lieb,Takahashi,Korepin} can be interpreted 
naturally at all energy scales in terms of the pseudohole and 
heavy-psepseudoparticle basis, but that such basis goes beyond 
that solution. We then establish that the 
generalized pseudoparticle perturbation theory, 
which is also naturally described in terms of the new operator 
algebra, can be used to study the low $(\omega -\omega_0)$ energy
excitations. Here $\omega_0$ represents the energy gaps
of particular finite-energy states II and non-LWS's and
non-HWS's which proven to dominate the finite-energy response functions
\cite{Carmelo96b,Carmelo96d}.

In contrast to the states I which at finite values of the
chemical potential and magnetic field span the low-energy
Hilbert subspace \cite{Carmelo96}, the LWS's and
HWS's II are characterized by finite occupancies of the
heavy-pseudoparticle bands, whose energy spectrum has a 
finite gap. In addition, our pseudohole and heavy-pseudoparticle 
operator basis generates the non-LWS's and non-HWS's of
the $\eta$-spin and spin algebras absent in the BA solution.
It goes beyond the BA solution \cite{Lieb,Takahashi,Korepin} because 
it i)- generates from a single vacuum all Hamiltonian eigenstates 
(including the above $\eta$-spin and spin non-LWS's and non-HWS's); 
ii)- provides relevant information on non-diagonal operators
\cite{Carmelo96b,Carmelo96d}; and
iii)- since the relevant finite-energy transitions involve
a small density of pseudoparticles, it allows the direct
study of the finite-energy transport properties
\cite{Carmelo96b,Carmelo96d}.

Importantly, we find that at {\it all} energy scales there is 
only zero-momentum pseudoparticle forward scattering in the Hubbard chain. 
Although this exact property holds only for {\it integrable}
systems, we believe that our pseudohole and heavy-pseudoparticle
operator description will provide the most suitable starting point for 
description of the unusual finite-frequency properties of both 
integrable and non-integrable electronic Luttinger liquids.
In the case of the latter liquids this could be achieved
by means of the bosonization of the pseudoparticle theory
\cite{Carmelo94}.

We fully define the pseudohole and heavy-pseudoparticle
statistics by evaluating their infinite-dimensional
statistical-interaction matrix \cite{Haldane91}. They
are neither fermions nor bosons, yet they obey an
anticommuting algebra. Furthermore, the introduction of 
topological-momentum-shift operators \cite{Carmelo96,Carmelo96c}
allows the use of a suitable second-quantization formalism. 

Since the Hamiltonian-eigenstate transitions involve only our
pseudohole, heavy-pseudoparticle, and topological-momentum-shift
operators, all operators acting onto the Hilbert space
can be written in terms of the former operators. 
However, the BA alone does not provide directly the expressions
of off-diagonal [in the energy eigenstate basis] part 
of the physical operators. Among the specific results we quote 
in this paper are the expressions in terms of pseudohole and
heavy-pseudoparticle operators for the Hamiltonian, momentum
operator, and six generators of the $\eta$-spin and spin $SO(4)$ 
algebra (yet four of these generators are off diagonal in 
the energy representation). Further, the algebra can provide 
diagonal part of all operators acting in the Hilbert space.
On the other hand, combining BA with the pseudoparticle
operator basis and the invariances
of the quantum liquid around particular values of 
momentum and energy provides important information
on correlation functions \cite{Carmelo96b,Carmelo96d}.
(This goes beyond BA because these functions are determined 
by the off-diagonal part of the physical operators.)
Our pseudohole and heavy-pseudoparticle operator representation
is the natural generalization of the low-energy pseudoparticle 
description which was studied in detail in Refs.
\cite{Carmelo90,Carmelo91,Carmelo92b,Carmelo92c,Carmelo93b,Carmelo94b}.

In Section II we introduce the pseudohole and heavy-pseudoparticle
operator description. In this operator representation all Hamiltonian
eigenstates have a simple Slater-determinant like form of pseudohole 
and heavy-pseudoparticle levels. 

In Section III we introduce the key concept of generalized ground
state (GGS) of which the GS is a particular case.
We also define the pseudohole and heavy-pseudoparticle 
statistics \cite{Haldane91}.

In Section IV we write the Hamiltonian in the pseudohole
and heavy-pseudoparticle operator basis. For this
we relate the rapidity numbers used
in the BA \cite{Takahashi} to the pseudomomentum-distribution 
operators and use this relation to write the Hamiltonian 
in terms of the pseudoparticle operators. We generalize
the rapidity operators for states out of the BA solution 
and show that the same rapidity eigenvalue describes a
$SU(2)$ $\eta$-spin tower or a $SU(2)$ spin tower of Hamiltonian
eigenstates.

In Sec. V we introduce the generalized pseudoparticle perturbation 
theory for Hilbert subspaces of arbitrary energy. This 
includes evaluation of the heavy-pseudoparticle energy gaps.
We discuss the effects of normal-ordering (relative to the 
suitable GGS or GS) and show that, in terms of the pseudoparticle 
operators, the normal-ordered Hamiltonian has in each subspace 
a simple form. This permits the introduction of a systematic, 
non-singular pseudoparticle perturbation theory valid at low $(\omega 
-\omega_0)$ energy and establishes directly the {\it universal} 
form of the Hamiltonian for integrable electronic quantum liquids. 

Finally, in Sec. VI we present the discussion and concluding 
remarks.

\section{THE GENERALIZED PSEUDOHOLE AND HEAVY-PSEUDOHOLE 
         OPERATOR BASIS}

We consider the Hubbard chain \cite{Lieb,Takahashi} in a magnetic 
field $H$ and chemical potential $\mu$ 

\begin{equation}
\hat{H} = \hat{H}_{SO(4)} + 2\mu\hat{S}^c_z + 
2\mu_0 H\hat{S}^s_z \, , 
\end{equation}
where 

\begin{equation}
\hat{H}_{SO(4)} = -t\sum_{j,\sigma}
\left[c_{j\sigma}^{\dag }c_{j+1\sigma} +
c_{j+1\sigma}^{\dag }c_{j\sigma}\right] +
U\sum_{j} [c_{j\uparrow}^{\dag }c_{j\uparrow}-{1\over 2}]
[c_{j\downarrow}^{\dag }c_{j\downarrow}-{1\over 2}] \, .
\end{equation}

In the absence of the chemical-potential and magnetic-field terms 
the Hamiltonian $(1)$ reduces to $(2)$ and has $SO(4) = SU(2) 
\otimes SU(2)/Z_2$ symmetry \cite{Heilmann,Yang89,Korepin,Ostlund}.
In equation $(1)$ we used the notations $\eta=S^c$ and $S=S^s$ 
for $\eta$-spin and spin, respectively. With this notation
the generators of the $\eta$-spin and spin $SO(4)$ algebra 
are \cite{Heilmann,Yang89,Korepin}, 

\begin{equation}
\hat{S}^c_z = -{1\over 2}[N_a - \sum_{\sigma}
\hat{N}_{\sigma }] \, , \hspace{1cm}
\hat{S}^c_- = \sum_{j} (-1)^j 
c_{j\uparrow}c_{j\downarrow} \, , \hspace{1cm}
\hat{S}^c_+=\sum_{j} (-1)^j c^{\dagger }_{j\downarrow}
c^{\dagger }_{j\uparrow} \, , 
\end{equation}
for $\eta$-spin and

\begin{equation}
\hat{S}^s_z = -{1\over 2}\sum_{\sigma}\sigma\hat{N}_{\sigma } 
\, , \hspace{1cm}
\hat{S}^s_- = \sum_{j}c^{\dagger }_{j\uparrow}c_{j\downarrow} 
\, , \hspace{1cm}
\hat{S}^s_+ = \sum_{j} c^{\dagger }_{j\downarrow}
c_{j\uparrow} \, , 
\end{equation}
for spin. Since $N_a$ is even, the operator ${\hat{S^c}}_z+{\hat{S^s}}_z$ 
has only integer eigenvalues and all half-odd integer representations of
$SU(2) \otimes SU(2)$ are projected out \cite{Heilmann,Yang89,Korepin}. 
This selection rule excludes transitions with 
$\Delta S^c_z+\Delta S^s_z$ half-odd integer.

In equations $(2)-(4)$ the operator $c_{j\sigma}^{\dagger}$ 
and $c_{j\sigma}$ creates and annihilates, respectively, one 
electron of spin projection $\sigma$ ($\sigma$ refers to the spin 
projections $\sigma =\uparrow\, ,\downarrow$ when used as an 
operator or function index and is given by $\sigma =\pm 1$ 
otherwise) at the site $j$, 

\begin{equation}
\hat{N}_{\sigma }=\sum_{j} 
c_{j\sigma }^{\dagger }c_{j\sigma } \, , 
\end{equation}
is the number operator for $\sigma$ spin-projection electrons, 
and $t$, $U$, and $\mu _0$ are the first-neighbor transfer 
integral, the onsite Coulomb repulsion, and the Bohr magneton, 
respectively. 

There are $N_{\uparrow}$ up-spin electrons and $N_{\downarrow}$ 
down-spin electrons in the chain of $N_a$ sites and
with lattice constant $a$ associated with the model
$(1)$. We use periodic boundary conditions and
consider $N_a$ to be even and when $N=N_a$ (half filling)
both $N_{\uparrow}$ and $N_{\downarrow}$ to be odd
and employ units such that $a=t=\mu_0=\hbar =1$. 
The dimensionless onsite repulsion is $u=U/4t$.
When $N_{\sigma }$ is odd the $U=0$ Fermi momenta are given by 
$k_{F\sigma }^{\pm }=\pm \left[k_{F\sigma } 
-{\pi\over {N_a}}\right]$ where

\begin{equation}
k_{F\sigma }={\pi N_{\sigma }\over {N_a}} \, .
\end{equation}
When $N_{\sigma }$ is even the Fermi momenta are given by
$k_{F\sigma}^{+} = k_{F\sigma}$ and $k_{F\sigma}^{-} = - 
\left[k_{F\sigma }-{2\pi\over {N_a}}\right]$ or by 
$k_{F\sigma}^{+} = \left[k_{F\sigma } -{2\pi\over {N_a}}\right]$ 
and $k_{F\sigma}^{-} = - k_{F\sigma }$. Often we can ignore 
the ${1\over {N_a}}$ corrections of these expressions
and consider $k_{F\sigma}^{\pm}\simeq\pm 
k_{F\sigma}=\pm \pi n_{\sigma }$ and $k_F=[k_{F\uparrow}+
k_{F\downarrow}]/2=\pi n/2$, where $n_{\sigma}=N_{\sigma}/N_a$ 
and $n=N/N_a$. We consider the electronic density $n=n_{\uparrow 
}+n_{\downarrow}$ and the spin density $m=n_{\uparrow}-n_{\downarrow}$. 

For finite values of both the magnetic field and chemical
potential the symmetry of the quantum problem is reduced to 
$U(1)\otimes U(1)$, with $\hat{S^c}_z$ and $\hat{S^s}_z$ 
commuting with $\hat{H}$ \cite{Carmelo96,Carmelo94b}. 
The eigenvalues ${S^c}_z$ and ${S^s}_z$ 
determine the different symmetries of the Hamiltonian
$(1)$. When $S^c_z\neq 0$ and $S^s_z\neq 0$ the symmetry is 
$U(1)\otimes U(1)$, for $S^c_z= 0$ and $S^s_z\neq 0$ (and 
$\mu =0$) it is $SU(2)\otimes U(1)$, when $S^c_z\neq 0$ 
and $S^s_z= 0$ it is $U(1)\otimes SU(2)$, and at $S^c_z= 0$ 
and $S^s_z= 0$ (and $\mu =0$) the Hamiltonian symmetry is $SO(4)$. 

There are four $U(1)\otimes U(1)$ sectors 
of parameter space which correspond to $S^c_z< 0$ and $S^s_z< 0$; 
$S^c_z< 0$ and $S^s_z> 0$; $S^c_z> 0$ and $S^s_z< 0$; and 
$S^c_z> 0$ and $S^s_z> 0$. We follow Ref. \cite{Carmelo95}
and refer these sectors by $(l_c,l_s)$ where

\begin{equation}
l_{\alpha} = {S^{\alpha }_z\over |S^{\alpha }_z|} \, .
\end{equation}
The sectors $(-1,-1)$; $(-1,1)$; $(1,-1)$; and $(1,1)$
refer to electronic densities and spin densities $0<n<1$ and  
$0<m<n$; $0<n<1$ and $-n<m<0$; $1<n<2$ and $0<m<(2-n)$;
and $1<n<2$ and $-(2-n)<m<0$, respectively.
  
There are two $(l_s)$ sectors of $SU(2)\otimes U(1)$ Hamiltonian 
symmetry [and two $(l_c)$ sectors of $U(1)\otimes SU(2)$ Hamiltonian 
symmetry] which correspond to $S^s_z< 0$ and $S^s_z> 0$ for
$l_s=-1$ and $l_s=1$, respectively, (and to $S^c_z< 0$ and $S^c_z> 0$ 
for $l_c=-1$ and $l_c=1$, respectively). There is one $SO(4)$ 
sector of parameter space [which is constituted only
by the $S^c_z= 0$ (and $\mu =0$) and $S^s_z= 0$ canonical
ensemble].

The related holon and spinon study of Ref. \cite{Essler} did not 
introduce the excitation operator generators (some of these were 
studied in Ref. \cite{Carmelo96}) and only refers and applies to 
states with eigenvalues ${S^c_z\over N_a}\sim 0$ and 
${S^s_z\over N_a}\sim 0$. In 
this paper we generate {\it all} Hamiltonian eigenstates from the 
pseudohole vacuum $|V\rangle $ introduced in Ref. \cite{Carmelo96}, 
which is the half-filled and zero-magnetic-field GS.
It is convenient for our complete operator description of
the quantum problem to consider explicitly all the above
sectors of parameter space.

In the case of integrable models of simple Abelian $U(1)$
symmetry the elementary excitations are generated by a single 
type of pseudoparticles (pseudoholes) \cite{Anto}. On the other hand,
in the present case of the Hubbard chain we have shown 
\cite{Carmelo96} that in terms of pseudoholes the description of the
states I involves four branches of $\alpha ,\beta$ pseudoholes, 
where $\alpha =c,s$ and $\beta=\pm {1\over 2}$. 
In the particular case of LWS's I and (or) HWS's I we have that the
$\alpha ,\beta$ pseudoholes are such that \cite{Carmelo96} 
$\beta={l_{\alpha}\over 2}$. Therefore, in each $(l_c,l_s)$ sector 
of parameter space of Hamiltonian symmetry $U(1)\otimes U(1)$ 
only the $c,{l_c\over 2}$ and $s,{l_s\over 2}$ 
pseudohole branches are involved in the description of the 
corresponding states I. 
However, the relation $\beta={l_{\alpha}\over 2}$ is not valid
for the general case considered in this paper, as we find below, yet 
the general description also involves the $\alpha,\beta$ pseudoholes 
introduced in Ref. \cite{Carmelo96}, which have creation operators 
$a^{\dag }_{q,\alpha,\beta}$. In addition, it
involves an infinite number of $\alpha,\gamma$ heavy-pseudoparticle 
branches, with creation operators $b^{\dag }_{q,\alpha,\gamma}$. 
Here the quantum numbers $\gamma $ are the positive integers ($\gamma
=1,2,...,\infty$). 

Elsewhere we will characterize the connection of the
pseudohole and heavy-pseudparticle quantum numbers to electronic
quantities in some limiting cases of physical interest.
For instance, in the $n\rightarrow 1$ and 
$S^c\rightarrow 0$ (or $m\rightarrow 0$ and $S^s\rightarrow 0$)
limit, $\alpha =c$ (or $\alpha =s$) becomes charge (or spin)
\cite{Carmelo94}, and the $c,\beta$ (or $s,\beta$)
pseudoholes are such that $\beta =S^c_z=\pm {1\over 2}$
(or $\beta =-S^s_z=\mp {1\over 2}$) \cite{Carmelo96}. [We 
emphasize that in that limit and in the particular case
we just considered $S^c_z=\pm {1\over 2}$
(or $S^s_z=\pm {1\over 2}$) denotes the $\eta $ spin
(or spin) projection of the individual $c,\beta$ (or $s,\beta$)
pseudohole. Therefore, it does not refer to the general definition  
associated with Eqs. $(3)$ and $(4)$ which defines 
the $\eta $ spin (or spin) projection of 
an Hamiltonian eigenstate.]
Moreover, for $U\rightarrow\infty$, the creation
of one $c,\gamma$ pseudoparticle ($\gamma >0$) is associated
with the creation of $D=\gamma$ doubly occupied sites. Note
that in that limit the Hamiltonian commutes with the
doubly-occupancy operator, i.e. $[\hat{H},\hat{D}]=0$.

The above operators obey an anticommuting algebra such that 

\begin{equation}
\{a^{\dag }_{q,\alpha,\beta},
a_{q',\alpha',\beta'}\}=\delta_{q,q'}\delta_{\alpha ,\alpha'}
\delta_{\beta,\beta'} \, ,
\end{equation}
and

\begin{equation}
\{b^{\dag }_{q,\alpha,\gamma},b_{q',\alpha'
,\gamma'}\}=\delta_{q,q'}\delta_{\alpha ,\alpha'}
\delta_{\gamma ,\gamma'} \, , 
\end{equation}
and all remaining anticommutators vanish.

As was found in Ref. \cite{Carmelo96}, the $\alpha ,\beta$
pseudoholes are the ``holes'' of the $\alpha $ pseudoparticles
considered in the low-energy representation introduced in Refs.
\cite{Carmelo92,Carmelo93,Carmelo94,Carmelo90} and
studied in more detail in Refs.
\cite{Carmelo91,Carmelo92b,Carmelo92c,Carmelo93b,Carmelo94b}.
In contrast to their ``holes'', the usual $\alpha $ 
pseudoparticles (of number $N_{\alpha}$ \cite{Carmelo92b,Carmelo94b})
are not labeled by the $\beta $ pseudohole
quantum number \cite{Carmelo96}. Moreover, there are no
$\beta $ bands, {\it i.e.} $\alpha, +{1\over 2}$ and 
$\alpha, -{1\over 2}$ pseudoholes occupy different 
pseudomomentum values in the same $\alpha $ band [as
the right-hand side (rhs) of Eq. $(23)$ below confirms].
For convenience we shall call the usual $\alpha $ pseudoparticles
\cite{Carmelo91,Carmelo92b,Carmelo92c,Carmelo93b,Carmelo94b}
and the corresponding $\alpha $ bands the $\alpha,0$ pseudoparticles
and $\alpha,0$ bands, respectively.

The description of the whole Hilbert space involves the set of 
$\alpha,\gamma$ pseudomomentum bands which are occupied by the 
$\alpha,\beta$ pseudoholes (or $\alpha,0$ pseudoparticles) for 
$\gamma =0$ and by the $\alpha,\gamma$ heavy pseudoparticles 
for $\gamma >0$. In particular,
the description of the non-LWS's and non-HWS's out 
of the BA requires the use of the $\alpha,\beta$ pseudoholes
instead of the usual $\alpha,0$ pseudoparticles. However, we
find in futures sections that both the $SO(4)$ Hamiltonian 
term $(2)$ and the BA rapidities are $\beta $ independent. Moreover, 
we find that the BA rapidities can be generalized to describe a
whole $SU(2)$ tower of $\eta $ spin or spin Hamiltonian
eigenstates. We emphasize that the BA solution refers
directly only to the LWS's (or HWS's) of these towers
\cite{Korepin}. In the case of such $\beta $-independent
quantities it is often convenient to use the 
$\alpha,0$-pseudoparticle representation rather than
the $\alpha ,\beta$-pseudohole representation. The 
$\alpha,0$ pseudoparticles have creation operators 
$b^{\dag }_{q,\alpha,0}$ and obey an anticommuting algebra such that 

\begin{equation}
\{b^{\dag }_{q,\alpha,0},b_{q',\alpha'
,0}\}=\delta_{q,q'}\delta_{\alpha ,\alpha'} \, , 
\end{equation}
and all remaining anticommutators vanish (including the ones
involving $\alpha,0$ pseudoparticles and heavy
pseudoparticles).

Often, when we refer $\alpha,\gamma$ pseudoparticles
(bands) we consider $\gamma =0,1,2,...$
whereas the notation $\alpha,\gamma$ {\it heavy} 
pseudoparticles refers only to $\gamma =1,2,...$.
The use of $\alpha,\beta$ pseudoholes and $\alpha,0$ 
pseudoparticles is {\it alternative} and in general we do not 
use them at the same time. While in the
characterization of all Hamiltonian eigenstates and in
studies of $\beta$-dependent operators the use of the 
$\alpha,\beta$-pseudohole and $\alpha,\gamma$-heavy-pseudoparticle 
representation {\it is required}, in studies involving $\beta $-independent
operators (as for instance in charge transport \cite{Carmelo96b})
it is often more convenient to use the $\alpha,\gamma$-pseudoparticle
representation (with $\gamma =0,1,2,3...$).

A crucial point is that the pseudohole and heavy-pseudoparticle 
numbers are constant for each Hamiltonian eigenstate.
The study of the non-LWS's outside the BA solution \cite{Korepin} requires 
the introduction of the generalized $\alpha,\beta$ pseudohole 
numbers. These are given by 

\begin{equation}
N^h_{c,\beta}={N^h_c\over 2}-\beta [N_a-N] \, ,  \hspace{1cm}
N^h_{s,\beta}={N^h_s\over 2}-\beta [N_{\uparrow}-
N_{\downarrow}] \, , 
\end{equation}
where 

\begin{equation}
N^h_c=N_a-N_{c, 0} \, ,  \hspace{2cm} N^h_s=N_{c, 0} - 
2\sum_{\gamma =0}^{\infty}N_{s,\gamma} \, ,
\end{equation}
and the total number of $\alpha $ pseudoholes is
$N^h_{\alpha} = \sum_{\beta}N^h_{\alpha,\beta}$.                

These important expressions are generalizations of the pseudohole
expressions of Ref. \cite{Carmelo96}, which refer only to
the $S^c$ and $S^s$ LWS's and HWS's I (which we call simply states 
I). It is convenient to introduce the numbers $N^z_{\alpha}$
such that

\begin{equation}
N^z_{\alpha} = S^{\alpha } - l_{\alpha}S^{\alpha }_z 
= S^{\alpha } - |S^{\alpha }_z| \, ,
\end{equation}
where for $S^{\alpha }_z\neq 0$ the parameter $l_{\alpha}=\pm 1$
is given in Eq. $(7)$. When dealing with energies and eigenvalues
instead of Hamiltonians and operators the numbers of Eq. $(13)$
together with the $\alpha ,\gamma$-pseudoparticle quantities 
and numbers constitute an alternative representation for the 
$\alpha ,\beta$-pseudohole and $\alpha ,\gamma$-heavy-pseudoparticle
quantities and numbers \cite{Carmelo96d}.
The knowledge of all the $N^h_{\alpha,\beta}$
pseudohole and $N_{\alpha,\gamma}$ heavy-pseudoparticle
($\gamma >0$) numbers provides the knowledge of all pseudoparticle
numbers $N_{\alpha,\gamma}$ ($\gamma =0,1,2,3...$) and
$N^z_{\alpha }$ numbers, and the inverse is also true. 
While the set of $N_{\alpha,\gamma}$ numbers
are directly given by the BA solution \cite{Takahashi}, 
the $\beta $-dependent pseudohole numbers $N^h_{\alpha,\beta}$ 
and the numbers $N^z_{\alpha }$ are not considered by that solution. 
In Appendix A we relate 
the pseudoparticle numbers $N_{\alpha,\gamma}$ with the
usual BA notation \cite{Takahashi}. On the other hand,
we find below that $\beta $ is the quantum number needed
to describe the Hamiltonian eigenstates out of the BA
solution. 

It follows from Eq. $(11)$ that the electron numbers are exclusive 
functions of the pseudohole numbers and read

\begin{equation}
N_{\uparrow} = {N_a\over 2} + \sum_{\beta}\beta[N^h_{c,\beta}
- N^h_{s,\beta}] \, , \hspace{1cm}
N_{\downarrow} = {N_a\over 2} + \sum_{\beta}\beta[N^h_{c,\beta}
+ N^h_{s,\beta}] \, . 
\end{equation}

As for the case of the Hilbert subspace spanned by the
states I, the pseudomomentum-number operators

\begin{equation}
\hat{N}^h_{\alpha,\beta}(q)=a^{\dag }_{q,\alpha 
,\beta}a_{q,\alpha ,\beta} \, , \hspace{1cm}
\hat{N}_{\alpha ,\gamma}(q)=
b^{\dag }_{q,\alpha ,\gamma}b_{q,\alpha ,\gamma} \, , 
\end{equation}
(with $\gamma =1,2,...$), and 

\begin{equation}
\hat{N}_{\alpha ,0}(q)\equiv 1
-\sum_{\beta}\hat{N}^h_{\alpha,\beta}(q) \, , 
\end{equation}
play a central role in the generalized theory. 
The operator $(16)$ has the following alternative representation
in terms of $\alpha,0$ pseudoparticle operators

\begin{equation}
\hat{N}_{\alpha ,0}(q)=
b^{\dag }_{q,\alpha ,0}b_{q,\alpha ,0} \, . 
\end{equation}
The number operators can be expressed in terms of
the pseudomomentum distributions as follows

\begin{equation}
\hat{N}^h_{\alpha ,\beta}=\sum_{q}\hat{N}^h_{\alpha ,\beta}(q) 
\, , \hspace{1cm}
\hat{N}_{\alpha,\gamma}=\sum_{q}\hat{N}_{\alpha 
,\gamma}(q) \, . 
\end{equation}

We emphasize that while all Hamiltonian eigenstates are
also eigenstates of the operator $\hat{N}^h_{\alpha,\beta}$
because these numbers are good quantum numbers, this does 
not hold true for the operator $\hat{N}^h_{\alpha,\beta}(q)$. 
We find below that only the states I are eigenstates of the 
latter operator, yet all Hamiltonian eigenstates are
also eigenstates of the pseudoparticle operators 
$\hat{N}_{\alpha ,\gamma}(q)$.

The $SO(4)$ operator generators, $\hat{S}^{\alpha }_z$ and 
$\hat{S}^{\alpha }_{\pm}$ [see Eqs. $(3)-(4)$], and the $\eta$-spin 
and spin numbers $S^{\alpha}$ and $N^z_{\alpha }$ [see Eq. $(13)$], 
are given by

\begin{equation}
\hat{S}^{\alpha }_z = \sum_{q,\beta}\beta
\hat{N}^h_{\alpha ,\beta}(q) \, ; \hspace{1cm}
{\hat{S}^{\alpha }}_{\pm} =\sum_{q}a^{\dag }_{q,\alpha,
\pm{1\over 2}}a_{q,\alpha,\mp{1\over 
2}} \, ,
\end{equation}
and 

\begin{equation}
S^{\alpha} = {1\over 2}[N^h_{\alpha}
-\sum_{\gamma =1}^{\infty}2\gamma N_{\alpha,\gamma}] \, ;
\hspace{1cm} N^z_{\alpha } =
{1\over 2}[\sum_{\beta}(1-2\beta l_{\alpha})\hat{N}^h_{\alpha ,\beta}
-\sum_{\gamma =1}^{\infty}2\gamma N_{\alpha,\gamma}] \, , 
\end{equation}
respectively. Note that the expressions for the off-diagonal
$SO(4)$ generators go beyond BA, which provides
the expressions only for diagonal operators. 
The application a suitable number of times of these 
operators onto the LWS's of the $\eta $-spin and spin 
algebras generates the non-LWS's towers of Hamiltonian
eigenstates outside the BA \cite{Korepin}. Therefore, their
off-diagonal form in the quantum number $\beta $, Eq. $(19)$,
reveals that in the present operator basis the construction 
of the states out of the BA requires the introduction 
of these pseudohole quantum numbers. This is confirmed by
the form of the Hamiltonian-eigenstate pseudohole
and heavy-pseudoparticle generators relative to the 
theory vacuum [see Eq. $(23)$ below].

The generalization of the LWS's BA total momentum expression
\cite{Essler,Takahashi}, which we denote by $P_{BA}$, 
to all Hamiltonian eigenstates, requires addition of an 
extra term associated with $\eta$ pairing 
\cite{Yang89,Korepin}. This leads to

\begin{equation}
P = P_{BA} + \pi[S^c + S^c_z] \, .
\end{equation}
When this expression provides values of the momentum such that
$|P|>\pi$, the total momentum is defined as the corresponding 
value at the first Brillouin zone.  
In operator form and for sub-canonical ensembles such that 
$S^{\alpha }_z\neq 0$ and with $l_{\alpha }$ given by Eq. $(7)$
we have that $N^h_{\alpha ,{-l_{\alpha}\over 2}}<
N^h_{\alpha ,{l_{\alpha}\over 2}}$
and the present representation leads to the
following simple expressions for the momentum operator 
$\hat{P}$ and its eigenvalue $P$

\begin{eqnarray}
\hat{P} & = & \sum_{q,\alpha}
\sum_{\gamma =0}^{\infty} qC_{\alpha,\gamma}
\hat{N}_{\alpha,\gamma}(q) +
\pi [\hat{N}^h_{c 
,{-l_c\over 2}}+\sum_{\gamma =1}\hat{N}_{c,\gamma}] \, ;
\nonumber \\
P & = & \sum_{q,\alpha}
\sum_{\gamma =0}^{\infty} qC_{\alpha,\gamma}
N_{\alpha,\gamma}(q) +
\pi\sum_{\gamma =1}\left[(1+\gamma)N_{c,\gamma} + N^z_c\right]
\, ,
\end{eqnarray}
respectively, where $C_{c,\gamma}=-1$ for $\gamma >0$ and 
$C_{\alpha,\gamma}=1$ otherwise. We emphasize that the momentum term, 
$\pi\sum_{\gamma =1}\left[(1+\gamma)N_{c,\gamma} + N^z_c\right]$,
is always a multiple of $\pm\pi$.

The pseudoparticle perturbation theory introduced in Refs. 
\cite{Carmelo92,Carmelo93,Carmelo94,Carmelo90,Carmelo91,Carmelo92b,Carmelo92c}
and developed in a suitable operator basis in Refs.
\cite{Carmelo94,Carmelo96,Carmelo94b} refers to the Hilbert 
subspace spanned by the Hamiltonian eigenstates I. Rather than 
holons and spinons \cite{Essler,Carmelo96,Faddeev}, at finite 
values of the magnetic field and chemical potential and at 
constant electronic numbers the low-energy excitations I are 
described by pseudoparticle-pseudohole processes relative to
the canonical-ensemble GS. The latter state, as well 
as all excited states I with the same electron numbers, are simple 
Slater determinants of pseudoparticle levels 
\cite{Carmelo94,Carmelo96,Carmelo95,Carmelo94b}. 

Although we find in Sec. III that the pseudoholes and heavy 
pseudoparticles obey generalized \cite{Haldane91} rather than 
Fermi or Bose statistics, one can use their formal 
anticommutation relations (together with suitable 
topological-momentum-shift operators introduced in Refs.
\cite{Carmelo96,Carmelo96c} and generalized for the
heavy pseudoparticles in Sec. III) to develop an extremely
simple second-quantized representation of {\it all}
$4^{N_a}$ orthonormal Hamiltonian eigenstates \cite{Korepin}. 
As for the above states I, in our generalized operator basis 
all these states are simple Slater determinants of 
pseudohole and heavy-pseudoparticle levels. 
They are of the form

\begin{equation}
|\psi;\{N^h_{\alpha ,\beta}\},\{N_{\alpha ,\gamma}\}\rangle = 
{1\over \sqrt{C}}\prod_{\alpha}A_{\alpha}\prod_{\gamma =1}^{\infty}
\left[[\hat{S}^{\alpha }_+]^{N^h_{\alpha ,{1\over 2}}}
\prod_{q,q'}a^{\dag }_{q,\alpha ,-{1\over 2}}
b^{\dag }_{q',\alpha ,\gamma}\right] 
|V\rangle \, ,
\end{equation}
where the generator $\hat{S}^{\alpha }_+$ is given in Eq.
$(19)$, $\{N^h_{\alpha ,\beta}\}$ and $\{N_{\alpha ,\gamma}\}$
are abbreviations for the sets 

\begin{equation}
\{N^h_{c,{1\over 2}},
N^h_{c,-{1\over 2}}, N^h_{s,{1\over 2}}, N^h_{s,-{1\over 2}}\}
\, ,
\end{equation}
and 

\begin{equation}
\{N_{c ,1},...,N_{c ,\infty}, N_{s ,1},..., N_{s 
,\infty}\} \, , 
\end{equation}
respectively, 

\begin{equation}
C=\prod_{\alpha}\Bigl({N^h_{\alpha }!\over
N^h_{\alpha ,-{1\over 2}}!}\Bigl) \, ; \hspace{1cm}
A_{\alpha} = \prod_{\beta}\Theta (N^h_{\alpha,\beta}-\sum_{\gamma =1}^{\infty}
\gamma N_{\alpha ,\gamma}) \, ,
\end{equation}
and $\Theta (x)=1$ for $x\geq 0$ and $\Theta (x)=0$ for $x<0$.
Note that $A_{\alpha}$ equals either $1$ or $0$. This is because in the 
present representation the Hamiltonian eigenstates are characterized by
pseudohole and heavy-pseudoparticle numbers such that 
$N^h_{\alpha,\beta}\geq\sum_{\gamma =1}^{\infty}\gamma N_{\alpha ,\gamma}$.
We note that application of the non-diagonal opertors of Eq. $(19)$
onto $\eta $ spin or spin LWS's or HWS's generates often states with 
numbers such that $N^h_{\alpha,\beta}<\sum_{\gamma =1}^{\infty}\gamma 
N_{\alpha ,\gamma}$. The presence of $A_{\alpha}$ (see Eq. $(26)$) 
in the rhs of Eq. $(23)$ (which is to be changed according to the changes
of the numbers $N^h_{\alpha ,\beta}$) then assures that application of 
the above generators in these LWS's or HWS's gives $0$.  

The existence of the half-filling Mott-Hubbard gap 
\cite{Lieb,Carmelo91}, which we denote here by $\Delta_{MH}$, 
implies that $S^c=S^c_z=0$ for chemical potentials $\mu$ such 
that $-{\Delta_{MH}\over 2}<\mu<{\Delta_{MH}\over 2}$.
Therefore, the $S^c=S^c_z=0$ and $S^s=S^s_z=0$ vacuum 
$|V\rangle $ of the rhs of Eq. $(23)$ can be associated 
with different chemical-potential values within the range 
$-{\Delta_{MH}\over 2}<\mu<{\Delta_{MH}\over 2}$. The vacuum 
chemical-potential value is choosen according to the suitable
reference level from which the energy of the pseudohole and
pseudoparticle seas 
of the states $(23)$ is to be measured. Therefore, we assume in
the definition of the vacuum of Eq. $(23)$ that 
$|V\rangle $ is such that $\mu =0$ for $S^c_z=0$ states 
but $\mu\rightarrow \pm{\Delta_{MH}\over 2}$ for
${S^c_z\over N_a}\rightarrow 0^{\mp}$.

The possible pseudohole and heavy-pseudoparticle numbers 
are constrained by Eqs. $(19)$ and $(20)$ and together with the
number of discrete pseudomomentum values in each band
leads to $4^{N_a}$ possible orthonormal Hamiltonian 
eigenstates of form $(23)$, in agreement with Ref.
\cite{Korepin}. The number of discrete pseudomomentum 
values in each ${\alpha,\gamma}$ band ($\alpha,\gamma$-band 
Fock-space dimension) is given by 

\begin{equation}
d_{F_{\alpha,\gamma}}=
N^h_{\alpha}+N_{\alpha,\gamma}-\sum_{\gamma'=1}^{\infty}
[\gamma + \gamma' - |\gamma - \gamma'|]N_{\alpha,\gamma'} \, . 
\end{equation}
In the above summations and products, the pseudomomentum takes on
values 

\begin{equation}
q_j={2\pi\over {N_a}}I_j^{\alpha,\gamma} \, ,
\end{equation}
where in contrast to the usual momentum, $I_j^{\alpha,\gamma}$ are 
consecutive integers or half-odd integers for ${\bar{N}}_{\alpha,\gamma}$ 
odd and even, respectively. Here 

\begin{equation}
{\bar{N}}_{c,0} = \sum_{\alpha}{N^h_{\alpha}\over 2} -
\sum_{\gamma =1}^{\infty}N_{c,\gamma} = {N_a\over 2} -
N_{s,0} - \sum_{\alpha ,\gamma =1}^{\infty}N_{\alpha,\gamma} \, ,
\end{equation}
and 

\begin{equation}
{\bar{N}}_{\alpha,\gamma}=
d_{F_{\alpha,\gamma}} \, , 
\end{equation}
for values of $\alpha ,\gamma $ other than $c, 0$.
It follows that for each $\alpha,\gamma$ 
band $q^{(-)}_{\alpha,\gamma}\leq q\leq q^{(+)}_{\alpha,\gamma}$, 
with the limits of the pseudo-Brillouin zones given by

\begin{equation}
q^{(\pm)}_{c,0}=\pm\pi[1-{1\over N_a}] \, ,
\end{equation}
for ${\bar{N}}_{c,0}$ even and 

\begin{equation}
q^{(+)}_{c,0}=\pi \, , \hspace{1cm}
q^{(-)}_{c,0}=-\pi[1-{2\over N_a}] \, ,
\end{equation}
or 

\begin{equation}
q^{(+)}_{c,0}=\pi[1-{2\over N_a}] \, , \hspace{1cm}
q^{(-)}_{c,0}=-\pi \, , 
\end{equation}
for ${\bar{N}}_{c,0}$ odd. On the other hand, the limits of the 
pseudo-Brillouin zones are simply given by 

\begin{equation}
q^{(\pm)}_{\alpha,\gamma}=\pm q_{\alpha,\gamma} = 
\pm{\pi\over N_a}[d_{F_{\alpha,\gamma}}-1] \, ,
\end{equation}
for all the remaining ${\alpha,\gamma}$ bands (other than the
$c,0$ band). 

The usual first-quantized BA wave function expression
for the Hamiltonian eigenstates has an involved form which
includes many permutations \cite{Lieb,Takahashi}.
That spatial wave function depends on the quantum numbers 
$I_j^{\alpha ,\gamma}$ of Eq. $(28)$ through infinite 
sets of complex numbers, which many authors call rapidities.
The expression of the spatial wave functions 
in terms of the quantum numbers $I_j^{\alpha ,\gamma}$ requires the 
solution of an infinite system of algebraic equations which 
we present in Sec. IV. These equations
define the real part of the BA rapidities as functions of the  
quantum numbers $I_j^{\alpha ,\gamma}$ \cite{Lieb,Takahashi}. Although 
the expression of the spatial wave function for the 
Hamiltonian eigenstates in terms of the quantum numbers 
$I_j^{\alpha ,\gamma}$ requires the solution of the above 
systems of equations, which constitutes a problem of 
considerable complexity, the description of these eigenstates 
in the basis of the above BA quantum numbers $I_j^{\alpha ,\gamma}$ 
that diagonalize the quantum liquid, Eq. $(23)$, is much simpler.
Historically, the Hamiltonian eigenstates were introduced
in terms of the spatial BA wave functions 
\cite{Bethe,Yang}. The diagonalization of the problem leads 
then to infinite systems of algebraic equations. These equations 
introduce the integer or half-odd integer quantum numbers 
$I_j^{\alpha ,\gamma}$ of Eq. $(28)$ which describe the 
Hamiltonian eigenstates. 

One of the principal advantages of the representation $(23)$
is that it permits the description of the Hamiltonian eigenstates 
in terms of the quantum numbers $I_j^{\alpha ,\gamma}$ 
and does not require the spatial wave-function representation. 
In the basis associated with these quantum numbers the description 
of these states does not involve the rapidity numbers.
Moreover, our algebraic representation $(23)$ refers
also to the Hamiltonian eigenstates out of the BA solution.

A $S^{\alpha}$ LWS is such that 

\begin{equation}
N^h_{\alpha ,{1\over 2}}=\sum_{\gamma 
=1}^{\infty}\gamma N_{\alpha ,\gamma} \, ,
\end{equation}
whereas for a $S^{\alpha}$ HWS we have that

\begin{equation}
N^h_{\alpha ,-{1\over 2}}=\sum_{\gamma =1}^{\infty}\gamma N_{\alpha 
,\gamma} \, , 
\end{equation}
with $N^h_{\alpha ,-{1\over 2}}\geq N^h_{\alpha 
,{1\over 2}}$ for a LWS and $N^h_{\alpha ,{1\over 2}}\geq N^h_{\alpha 
,-{1\over 2}}$ for a HWS. 

For the particular case of states I, we have that
$N_{\alpha ,\gamma}=0$ for all $\gamma =1,2,...$ branches in 
the former equations. Therefore, the states I are
characterized by zero occupancies of the heavy-pseudoparticle
bands $\alpha ,\gamma$ (with $\gamma >0$). It then
follows from the general Eqs. $(35)$ and $(36)$
that a $S^{\alpha}$ LWS I is such that 

\begin{equation}
N^h_{\alpha ,{1\over 2}} = 0 \, ,
\end{equation}
whereas for a $S^{\alpha}$ HWS I we have that

\begin{equation}
N^h_{\alpha ,-{1\over 2}}= 0 \, , 
\end{equation}
with $N^h_{\alpha ,-{1\over 2}}\geq 0$ for a LWS I and 
$N^h_{\alpha ,{1\over 2}}\geq 0$ for a HWS I. This together with
Eq. $(23)$ leads to the simple Slater-determinant form found
for the states I in Ref. \cite{Carmelo96}. These 
results also confirm that in the case of states
I the $\alpha ,\beta$ pseudoholes are such that
$\beta ={l_{\alpha }\over 2}$ but that this equality
does not hold true in the general case. 
While in the case of the states I the Slater
determinant $(23)$ either involves $\alpha, +{1\over 2}$
or $\alpha, -{1\over 2}$ pseudoholes only, in the case
of the non-LWS's and non-HWS's or in the case of
states II (i.e. LWS's or HWS's containing $\alpha ,\gamma$ heavy 
pseudoparticles) the state $(23)$ has both $\alpha, +{1\over 2}$
and $\alpha, -{1\over 2}$ pseudohole occupancies. 

In the case of the non-LWS's and non-HWS's
out of BA the relative number of $\alpha ,{1\over 2}$ and
$\alpha ,-{1\over 2}$ pseudoholes is determined
by the values of $S^{\alpha }$ and $S^{\alpha }_z$
through Eqs. $(19)$, $(20)$, and $(39)$. The pseudohole numbers 
which characterize these states are such that

\begin{equation}
N^h_{\alpha}-\sum_{\gamma =1}^{\infty}2\gamma 
N_{\alpha ,\gamma}>|N^h_{\alpha ,{1\over 2}}-
N^h_{\alpha ,-{1\over 2}}| \, .
\end{equation}
They can be generated from LWS's (or HWS's) by acting onto the latter the 
off-diagonal generator ${\hat{S}}^{\alpha }_+$ (or ${\hat{S}}^{\alpha }_-$)
the required number of times. According to Eq. $(19)$, in the present basis 
these off-diagonal generators just produce $\beta $-flip processes.
In the case of the LWS's II the $\beta $ 
flips generated by these operators create a number $\gamma $ of 
$\alpha ,{1\over 2}$ pseudoholes for each $\alpha ,\gamma$ heavy 
pseudoparticle, which implies these states are inside BA. 
Following Eqs. $(19)-(20)$ this is required for the LWS 
condition $S^{\alpha }=-S^{\alpha }_z$. Note that both for the
non-LWS's and non-HWS's out of the BA and for the LWS's II
and HWS's II there is no relation between the quantity
$l_{\alpha }$ defined in Eq. $(7)$ and the pseudohole quantum 
number $\beta $.

In the case of non-LWS's and non-HWS's the wave function $(23)$ 
uses as starting state a LWS. We then call it for LWS general 
representation. The general state $(23)$ can be rewritten, 
alternatively, in terms of a starting HWS (HWS 
general representation). This just requires replacing in the 
rhs of Eq. $(23)$ $\pm {1\over 2}$ by $\mp {1\over 2}$
and the generator $\hat{S}^{\alpha }_+$ by $\hat{S}^{\alpha }_-$. In 
the corresponding HWS general representation the off-diagonal generators
create new states from a reference HWS. One can also use a
representation where the starting state is a $S^c$ (or $S^s$)
HWS and a $S^s$ (or $S^c$) LHS. However, the four possible general 
representations describe the same states. 

The form of the pseudohole and heavy-pseudoparticle generators 
of the rhs of Eq. $(23)$ confirms that the corresponding Hamiltonian 
eigenstates are not in general eigenstates of the pseudohole 
operator $\hat{N}^h_{\alpha,\beta}(q)$. This is because due
to the transformation laws of the off-diagonal $\eta$-spin and spin 
generators present in the rhs of Eq. $(23)$ a Hamiltonian eigenstate
is, in general, a mixture of different $\beta =-{1\over 2}$
and $\beta =+{1\over 2}$ pseudomomentum distributions.   
The exception are the states I which have only either $\alpha,{1\over 2}$ 
or $\alpha,-{1\over 2}$ pseudohole occupancy. 

\section{GENERALIZED GROUND STATES AND STATISTICS}

The conservation of the pseudohole and heavy-pseudoparticle
numbers permits dividing the Hilbert space into
subspaces spanned by the set of Hamiltonian 
eigenstates $(23)$ with the same $\{N^h_{\alpha ,\beta}\}$
and $\{N_{\alpha ,\gamma}\}$ numbers, which
correspond to the same sub-canonical ensemble.
Obviously, a canonical ensemble (with constant electron 
numbers and thus with constant values of $N^h_{\alpha 
,{1\over 2}}-N^h_{\alpha ,-{1\over 2}}$ for both $\alpha=c,s$) 
is usually realized by several $\{N^h_{\alpha ,\beta}\}$,
$\{N_{\alpha ,\gamma}\}$ sub-canonical ensembles. 
Let us introduce the generalized ground state (GGS) as the 
Hamiltonian eigenstate(s) $(23)$ of lowest energy in each
Hilbert subpace. We find it is (they are) of the form

\begin{equation}
|GGS;\{N^h_{\alpha ,\beta}\},\{N_{\alpha ,\gamma}\}\rangle = 
{1\over \sqrt{C}}\prod_{\alpha}A_{\alpha}\prod_{\gamma =1}^{\infty}
\left[[\hat{S}^{\alpha }_+]^{N^h_{\alpha ,{1\over 2}}}
\prod_{q=q_{\alpha,0}^{(-)}}^{{\bar{q}}_{F\alpha,0}^{(-)}} 
\prod_{q={\bar{q}}_{F\alpha,0}^{(+)}}^{q_{\alpha,0}^{(+)}} 
\prod_{q'=q_{c ,\gamma}^{(-)}}^{q_{Fc,\gamma}^{(-)}}
\prod_{q'=q_{Fc ,\gamma}^{(+)}}^{q_{c,\gamma}^{(+)}}
\prod_{q''=q_{Fs ,\gamma}^{(-)}}^{q_{Fs,\gamma}^{(+)}}
a^{\dag }_{q,\alpha ,-{1\over 2}}
b^{\dag }_{q',c ,\gamma}
b^{\dag }_{q'',s ,\gamma}
\right]|V\rangle \, ,
\end{equation}
where

\begin{equation}
{\bar{q}}_{F\alpha,\gamma}^{(\pm)}=q_{F\alpha,
\gamma}^{(\pm)}\pm C_{\alpha ,\gamma}{2\pi\over {N_a}} \, ,
\end{equation}
and when (i) $\alpha =c$, $\gamma >0$, and $N_{c,\gamma}$ is even or 
(ii) $\alpha =s$ or $\alpha =c$ and $\gamma =0$ and 
$N_{\alpha,\gamma}$ is odd (even) and $I_j^{\alpha,\gamma}$ are 
integers (half-odd integers) the pseudo-Fermi points are symmetric 
and given by 

\begin{equation}
q_{F\alpha,\gamma}^{(\pm)}=\pm [q_{F\alpha,\gamma}
- C_{\alpha ,\gamma}{\pi\over {N_a}}] \, , 
\end{equation}
where 

\begin{eqnarray}
q_{Fc,\gamma} & = & {\pi[d_{c,\gamma}-N_{c,\gamma}]\over N_a} \, ;
\hspace{1cm} \gamma >0 \, , \nonumber \\
q_{Fc,0} & = & {\pi N_{c,0}\over N_a} \, , \nonumber \\ 
q_{Fs,\gamma} & = & {\pi N_{s,\gamma}\over N_a} \, .
\end{eqnarray}
On the other hand, when (i) $\alpha =c$, $\gamma >0$, and 
$N_{c,\gamma}$ is odd or (ii) $\alpha =s$ or $\alpha =c$ and
$\gamma =0$ and $N_{\alpha,\gamma}$ is odd (even) and 
$I_j^{\alpha,\gamma}$ are half-odd integers (integers) we have that 

\begin{equation}
q_{F\alpha,\gamma}^{(+)}=q_{F\alpha,\gamma} \, , \hspace{1cm}
q_{F\alpha,\gamma}^{(-)} =-[q_{F\alpha,\gamma}-
C_{\alpha ,\gamma}{2\pi\over N_a}] \, , 
\end{equation}
or 

\begin{equation}
q_{F\alpha,\gamma}^{(+)}=q_{F\alpha,\gamma}-
C_{\alpha ,\gamma}{2\pi\over {N_a}} 
\, , \hspace{1cm}
q_{F\alpha,\gamma}^{(-)}= -q_{F\alpha,\gamma} \, . 
\end{equation}

The GS associated with a given canonical ensemble is always a state 
I of the form presented in Ref. \cite{Carmelo96}, which is
a particular case of the general GGS expression $(40)$.
It is useful to denote the GS's by $|GS;S^c_z,S^s_z\rangle$,
where $S^c_z$ and $S^s_z$ are the eigenvalues of the
corresponding canonical ensemble.
The GS expression associated with the $(l_c,l_s)$ 
sector of Hamiltonian symmetry $U(1)\otimes U(1)$ reads 

\begin{equation}
|GS;S^c_z,S^s_z\rangle = 
\prod_{q=q_{c,0}^{(-)}}^{{\bar{q}}_{Fc,0}^{(-)}} 
\prod_{q={\bar{q}}_{Fc,0}^{(+)}}^{q_{c,0}^{(+)}} 
a^{\dag }_{q,c,{l_c\over 2}} 
\prod_{q=q_{s,0}^{(-)}}^{{\bar{q}}_{Fs,0}^{(-)}} 
\prod_{q={\bar{q}}_{Fs,0}^{(+)}}^{q_{s,0}^{(+)}} 
a^{\dag }_{q,s,{l_s\over 2}} 
|V \rangle \, .
\end{equation}

In all sectors of Hamiltonian symmetry there are states I. In the
particular case of the $SO(4)$ zero-chemical potential
and zero-magnetic field canonical ensemble there is only one state 
I, which is the pseudohole and heavy-pseudoparticle vacuum, 
$|V \rangle$, which is nothing but the $SO(4)$ GS
\cite{Carmelo96}. The same applies to the sectors 
of Hamiltonian symmetry $SU(2)\otimes U(1)$ and $U(1)\otimes SU(2)$, 
the GS being always a state I. (In addition, in these
sectors there is a large number of excited states I.)

In the case of the two $(l_s)$ $SU(2)\otimes U(1)$ 
sectors the GS is both a LWS and HWS of the $\eta$-spin algebra. 
Therefore, it is empty of $c$ pseudoholes and reads

\begin{equation}
|GS;0,S^s_z\rangle = 
\prod_{q=q_{s,0}^{(-)}}^{{\bar{q}}_{Fs,0}^{(-)}} 
\prod_{q={\bar{q}}_{Fs,0}^{(+)}}^{q_{s,0}^{(+)}} 
a^{\dag }_{q,s,{l_s\over 2}} 
|V\rangle \, .
\end{equation}

In the case of the $(l_c)$ $U(1)\otimes SU(2)$ sector 
the GS is both a LWS and a HWS of the spin algebra
and is empty of $s$ pseudoholes. It reads

\begin{equation}
|GS;S^c_z,0\rangle = 
\prod_{q=q_{c,0}^{(-)}}^{{\bar{q}}_{Fc,0}^{(-)}} 
\prod_{q={\bar{q}}_{Fc,0}^{(+)}}^{q_{c,0}^{(+)}} 
a^{\dag }_{q,c,{l_c\over 2}} |V\rangle \, .
\end{equation}

Finally, the $S^c=S^c_z=0$ (and $\mu =0$) and $S^s=S^s_z=0$ $SO(4)$ 
ground state is, at the same time, a LWS and HWS of both the $\eta$-spin 
and spin algebras, ie following the notation of Refs. 
\cite{Carmelo96,Carmelo95} 
it is a [LWS,LWS], a [LWS,HWS], a [HWS,LWS], and a 
[HWS,HWS]. Therefore, it is empty of both $c$ and $s$
pseudoholes and is the vacuum of the pseudohole and
heavy-pseudoparticle theory. All the remaining states I can be 
described by particular cases of the general expression
$(23)$ which are Slater determinants of $\alpha ,\beta$ 
pseudoholes and $\alpha,\gamma$ heavy pseudoparticles \cite{Carmelo96}.

The number of Hamiltonian eigenstates associated with a
sub-canonical ensemble and thus characterized by the same 
$\{N^h_{\alpha ,\beta}\}$ and $\{N_{\alpha ,\gamma}\}$ numbers 
is given by

\begin{equation}
n\left(\{N^h_{\alpha ,\beta}\},\{N_{\alpha ,\gamma}\}\right) =
\prod_{\alpha =c,s}\left[
\begin{array}{c}
d_{F\alpha ,0}\\
N^h_{\alpha ,+{1\over2}}+N^h_{\alpha ,-{1\over2}}
\end{array}
\right]
\prod_{\gamma =1}^{\infty}
\left[
\begin{array}{c}
d_{F\alpha ,\gamma}\\
N_{\alpha ,\gamma} 
\end{array}
\right] = 
\prod_{\alpha =c,s}\left[
\begin{array}{c}
d_{F\alpha ,0}\\
N^h_{\alpha}
\end{array}
\right]
\prod_{\gamma =1}^{\infty}
\left[
\begin{array}{c}
d_{F\alpha ,\gamma}\\
N_{\alpha ,\gamma} 
\end{array}
\right] \, , 
\end{equation}
which in terms of pseudoparticle numbers only can be
rewritten simply as

\begin{equation}
\prod_{\alpha =c,s}
\prod_{\gamma =0}^{\infty}
\left[
\begin{array}{c}
d_{F\alpha ,\gamma}\\
N_{\alpha ,\gamma} 
\end{array}
\right] \, .
\end{equation}
The square brackets in the above
equations refer to the usual combinatoric coefficients.

To classify the statistics of our pseudoholes and 
heavy pseudoparticles according to the generalized Pauli principle
of Ref. \cite{Haldane91}, we introduce the pseudohole and
heavy-pseudoparticle dimensions 

\begin{equation}
d^h_{\alpha}\equiv 1+d_{F_{\alpha,0}}
-N^h_{\alpha} \, ,  \hspace{1cm}
d_{\alpha,\gamma}\equiv 1+d_{F_{\alpha,\gamma}}
-N_{\alpha,\gamma} \, , 
\end{equation}
respectively. An Hamiltonian eigenstate transition producing changes 
$\Delta N^h_{\alpha}$ and $\Delta N_{\alpha,\gamma}$ in
the pseudohole and heavy pseudoparticle numbers, respectively,
leads then to the following changes in the corresponding
dimensions

\begin{equation}
\Delta d^h_{\alpha}=-\sum_{\alpha'}g_{\alpha,\alpha'}
\Delta N^h_{\alpha'}-\sum_{\alpha'}\sum_{\gamma'=1}^{\infty}
g_{\alpha;\alpha',\gamma'}\Delta N_{\alpha',\gamma'} \, ,
\end{equation}
and

\begin{equation}
\Delta d_{\alpha,\gamma}=-\sum_{\alpha'}g_{\alpha,\gamma;\alpha'}
\Delta N^h_{\alpha'}-\sum_{\alpha'}\sum_{\gamma'=1}^{\infty}
g_{\alpha,\gamma;\alpha',\gamma'}\Delta N_{\alpha',\gamma'} \, ,
\end{equation}
where the statistical-interaction matrix has infinite
dimension and reads 

\begin{equation}
g_{c,\alpha}=\delta_{c,\alpha} \, , \hspace{1cm}
g_{c;\alpha,\gamma}=0 \, , 
\end{equation}

\begin{equation}
g_{s,\alpha}={1\over 2} \, , \hspace{1cm}
g_{s;\alpha,\gamma}=\delta_{s,\alpha} \, ,
\end{equation}

\begin{equation}
g_{\alpha,\gamma;\alpha'}=-\delta_{\alpha,\alpha'} \, , \hspace{1cm}
g_{\alpha,\gamma;\alpha',\gamma'}=\delta_{\alpha,\alpha'}
\Bigl(\gamma + \gamma' - |\gamma - \gamma'|\Bigl) \, . 
\end{equation}
This fully defines the statistics, the pseudoholes and 
heavy pseudoparticles being neither fermions nor bosons 
\cite{Haldane91}.

The simple form of the GGS expression $(40)$, GS expressions 
$(46)-(48)$, and general-Hamiltonian eigenstates $(23)$ has a 
deep physical meaning. It reveals that in the present
basis these eigenstates of the many-electron quantum problem 
are ``non-interacting'' states of simple 
Slater-determinant form. However, that the
numbers $I_j^{\alpha ,\gamma}$ of the rhs of Eq. $(28)$ can 
be integers or half-odd integers for different Hamiltonian 
eigenstates, makes the problem much more involved than a simple 
non-interacting case. This change in the integer or half-odd
integer character of 
some of the numbers $I_j^{\alpha ,\gamma}$ of two
states, shifts {\it all} the occupied pseudomomenta
and leads to the orthogonal catastrophes \cite{Anderson}. 
These shifts are associated with topological excitations
which are generated by the topolological-momentum shift
operators introduced in Refs. \cite{Carmelo96,Carmelo96c}
for the particular case of the states I.

Let us introduce the $\alpha,0$ and 
$\alpha,\gamma$ topological-momentum-shift operators (for the states 
I see Refs. \cite{Carmelo96,Carmelo96c})
which generate topological 
excitations associated with a shift of all the pseudomomenta of
the corresponding band by $\pm\frac{\pi}{N_a}$. 
Creation or annihilation of odd numbers of pseudoholes or 
heavy-pseudoparticles {\it always} requires the occurrence of such 
topological-momentum shifts. Therefore, in addition to the pseudohole
and heavy-pseudoparticle operators our second-quantized algebra
also includes topological-momentum shift operators. 
Their unitary generators are such that

\begin{equation}
U^{\pm 1}_{\alpha,0}a^{\dag }_{q,\alpha,\beta}=
a^{\dag }_{q\pm {\pi\over N_a},\alpha,\beta}U^{\pm 1}_{\alpha} \, ,
\end{equation}
and

\begin{equation}
U^{\pm 1}_{\alpha,\gamma}b^{\dag }_{q,\alpha,\gamma}=
b^{\dag }_{q\pm {\pi\over N_a},\alpha,\gamma}U^{\pm 1}_{\alpha,
\gamma} \, , 
\end{equation}
and read 

\begin{equation}
U^{\pm 1}_{\alpha,0}=\exp[-\sum_{q,\beta}
a^{\dag }_{q\pm {\pi\over N_a},\alpha,\beta}a_{q,\alpha,\beta}] \, , 
\end{equation}
and 

\begin{equation}
U^{\pm 1}_{\alpha,\gamma}=\exp[\sum_{q}b^{\dag }_{q\mp 
{\pi\over N_a},\alpha,\gamma}b_{q,\alpha,\gamma}] \, . 
\end{equation}

As we discuss in future sections, a remarkable property is 
that the transition between a GS $(46)$ and {\it any} eigenstate 
$(23)$ can be separated into two types of excitations: (a) a 
topological GS -- GGS transition which involves the creation or 
annihilation of pseudoholes and (or) heavy pseudoparticles as 
well as the occurrence of topological momentum shifts and (b) a {\it 
Landau-liquid} excitation \cite{Carmelo94,Carmelo96} associated 
with pseudoparticle - pseudohole processes relative to the GGS. 

The topological transitions (a) 
are basically superpositions of three kinds of elementary 
transitions: (i) GS - GS transitions involving changes in
the $\sigma $ electron numbers by one, (ii) single $\beta$ 
flip processes of $\alpha,\beta$ pseudoholes 
which lead to non-LWS's and non-HWS's outside the BA,
and (iii) creation of single $\alpha,\gamma$ pseudoparticles at 
constant values of $S^c$, $S^c_z$, $S^s$, and $S^s_z$. 
While the transitions (i) are gapless, the elementary
excitations (ii) and (iii) require a finite amount of 
energy. 

The generators of the excitations (i) were 
studied in Ref. \cite{Carmelo96} and the ones of (ii) involve
only topological momentum shifts and $\beta $ flips (which describe either 
creation of electron pairs or spin flip processes). Consider 
as an example of a transition (iii) the creation of 
one $\alpha,\gamma $ pseudoparticle at the pseudo-Fermi points,
$q=q^{(\pm)}_{F\alpha,\gamma}$. 
(We examine here explicitly topological momentum shifts of occupied bands 
only, {\it i.e.} these which generate momentum.) Let us introduce the notation 
$-\alpha $ such that $-c=s$ and $-s=c$. When $\gamma >0$ 
is even (or odd) the generator reads $U^{\pm 1}_{\alpha,0}
G_{\alpha,\gamma}$ (or $G_{\alpha,\gamma}U^{\pm 1}_{-\alpha,0}$), where the 
topological-momentum-shift operator was defined above and

\begin{equation}
G_{\alpha,\gamma}=[\hat{S}^{\alpha}_+]^{\gamma}
\prod_{q=q_{F\alpha,0}^{(-)}}^{q_{F\alpha,0}^{(-)}+
{2\pi\over N_a}(\gamma -1)}\prod_{q_{F\alpha,0}^{(+)}-
{2\pi\over N_a}(\gamma -1)}^{q_{F\alpha,0}^{(+)}} 
a^{\dag }_{q,\alpha,-{1\over 2}}
b^{\dag }_{q^{(\pm)}_{F\alpha,\gamma},\alpha,\gamma} \, .
\end{equation}
Therefore, such transitions involve one $\alpha,0$ or $-\alpha,0$
topological momentum shift, the creation of a number $\gamma$ of
$\alpha,{1\over 2}$ pseudoholes and $\gamma$ of $\alpha,-{1\over 2}$ pseudoholes, 
and the creation of one $\alpha,\gamma$ pseudoparticle at
one of the pseudo-Fermi points. (This can be considered as the antiparticle of
the ``cloud'' of $2\gamma $ associated pseudoholes.) 

\section{PSEUDOPARTICLE HAMILTONIAN AND BA RAPIDITIES}

For simplicity let us denote the general Hamiltonian
eigenstates $|\psi;\{N^h_{\alpha ,\beta}\},\{N_{\alpha 
,\gamma}\}\rangle$ by $|\psi\rangle$ [and the GGS's 
$(40)$ by $|GGS\rangle$.] These states are
also eigenstates of the $\alpha,\gamma$ 
pseudomomentum-distribution operators $(15)$ such that

\begin{equation}
\hat{N}_{\alpha ,\gamma}(q)
|\psi\rangle  = N_{\alpha,\gamma}(q)|\psi\rangle \, ,
\end{equation}
where $N_{\alpha ,\gamma}(q)$ represents the 
eigenvalue of the operator $(15)$, which is given by $1$ and $0$ 
for occupied and empty values of $q$, respectively. 

It follows from the form of the state-generator of the rhs of
Eq. $(23)$ that all $2S^{\alpha }+1$ Hamiltonian eigenstates 
constructed from the same $S^{\alpha }=-S^{\alpha }_z$ LWS by applying 
onto it $1,2,...,2S^{\alpha }$ number of times the off-diagonal operator 
${\hat{S}}_+^{\alpha }$ $(19)$ have the same $N_{\alpha,\gamma}(q)$
pseudomomentum distribution.
On the other hand, the pseudohole numbers $N^h_{\alpha,-{1\over 2}}$
and $N^h_{\alpha,+{1\over 2}}$ are different for each of  
the $2S^{\alpha }+1$ states of such $S^{\alpha }$ tower.
These numbers change from $N^h_{\alpha,-{1\over 2}}=N^h_{\alpha}-
\sum_{\gamma =1}^{\infty}\gamma N_{\alpha,\gamma}$
and $N^h_{\alpha,+{1\over 2}}=\sum_{\gamma =1}^{\infty}\gamma N_{\alpha,\gamma}$ 
for the initial LWS to $N^h_{\alpha,-{1\over 2}}=\sum_{\gamma =1}^{\infty}\gamma 
N_{\alpha,\gamma}$ and $N^h_{\alpha,+{1\over 2}}=N^h_{\alpha}-
\sum_{\gamma =1}^{\infty}\gamma N_{\alpha,\gamma}$ for the corresponding HWS, 
where $N^h_{\alpha}$ is given in Eq. $(12)$. 

As in the case of the spatial wave functions, the expression of 
the energy in terms of the quantum numbers $I_j^{\alpha ,\gamma}$
of Eq. $(28)$ involves the BA rapidities. These are complex functions
of the numbers $I_j^{\alpha ,\gamma}$. Following our pseudomomentum
definition, Eq. $(28)$, in our description they are functions of
the pseudomomentum $q$. The real part of the BA rapidities, which we 
denote by $R_{\alpha,\gamma}(q)$, are in the present 
basis eigenvalues of corresponding {\it rapidity operators},
$\hat{R}_{\alpha,\gamma}(q)$. If we combine our operator 
representation with the properties of the BA solution, it is 
straightforward to confirm from the relation between the 
rapidity real part, $R_{\alpha,\gamma} (q)$, and the pseudomomentum 
distributions of Eq. $(62)$ that these real rapidity-function parts are   
{\it eigenvalues} of the rapidity operators 
$\hat{R}_{\alpha}(q)$. These rapidity operators are 
exclusive functions of the pseudoparticle operators
$\hat{N}_{\alpha ,\gamma}(q)$. It follows that all Hamiltonian
eigenstates are also eigenstates of the rapidity
operators $\hat{R}_{\alpha,\gamma}(q)$ such that

\begin{equation}
\hat{R}_{\alpha,\gamma}(q)|\psi\rangle = 
R_{\alpha,\gamma}(q)|\psi\rangle \, ,
\end{equation}
where for the particular case of $R_{c,0}(q)$ we also consider
the associate rapidity $K(q)$ which is defined by the following 
equation

\begin{equation}
R_{c,0}(q) = {\sin K(q)\over u} \, .
\end{equation}
This notation refers to the spectral parameter ${\sin K(q)\over u}$
which appears often in operator form. The rapidity $K(q)$ is the 
eigenvalue of the rapidity operator $\hat{K}(q)$ such that

\begin{equation}
\hat{K}(q)|\psi\rangle = K(q)|\psi\rangle \, .
\end{equation}
In Appendix A we present the relation of our rapidity eigenvalues
to the real part of the BA rapidity numbers \cite{Takahashi}.

The real part of the BA rapidities, which in our basis are the
eigenvalues $R_{\alpha,\gamma}(q)$ [and $K(q)$], are 
fully determined by the pseudoparticle distributions
$N_{\alpha,\gamma}(q)$ through a system of coupled integral 
equations. The standard treatment of the BA 
solutions is very lengthy, {\it e.g.}, for each state we have to 
{\it rewrite} a new set of equations. This is because the 
eigenvalues $R_{\alpha,\gamma} (q)$ of the rapidity 
operators $\hat{R}_{\alpha,\gamma}(q)$ are different for each state 
$|\psi\rangle$ $(23)$. 

However, within the present operator description,
we can introduce a single set of infinite
general equations which apply to {\it all} LWS's.
In addition, our operator representation allows the generalization 
of the BA rapidities to the non-LWS's out of BA. It follows
that the above general equations can be generalized
to {\it all} Hamiltonian eigenstates. 
Each of these equations defines one of the rapidity 
operators $\hat{R}_{\alpha,\gamma}(q)$ in terms of a $q$, 
$\alpha$, and $\gamma$ summation containing functionals of the 
rapidity operators and pseudomomentum-distribution operators. 
This system of coupled equations has a unique solution which
defines the expressions for the rapidity operators in 
terms of the pseudomomentum-distribution 
operators. In terms of the corresponding eigenvalues
the system of coupled integral equation reads

\begin{eqnarray}
K(q) & = & q - {2\over N_a}\sum_{\gamma =0}\sum_{q'}N_{s,\gamma}(q')
\tan^{-1}\Bigl({R_{c,0}(q)-R_{s,\gamma}(q')\over 1+\gamma}\Bigr)
\nonumber \\
& - & {2\over N_a}\sum_{\gamma =1}\sum_{q'}N_{c,\gamma}(q')
\tan^{-1}\Bigl({R_{c,0}(q)-R_{c,\gamma}(q')
\over \gamma}\Bigr) \, ,
\end{eqnarray}

\begin{eqnarray}
2Re \sin^{-1}\Bigl(u(R_{c,\gamma}(q)-i\gamma)\Bigr) 
& = & q + {2\over N_a}\sum_{q'}N_{c,0}(q')
\tan^{-1}\Bigl({R_{c,\gamma}(q)-R_{c,0}(q')\over \gamma}\Bigr)
\nonumber \\
& + & {1\over N_a}\sum_{\gamma' =1}\sum_{q'}N_{c,\gamma'}(q')
\Theta_{\gamma,\gamma'}\Bigl(R_{c,\gamma}(q)-R_{c,\gamma'}(q')\Bigr) \, ,
\end{eqnarray}
and 

\begin{eqnarray}
0 & = & q - {2\over N_a}\sum_{q'}N_{c,0}(q')
\tan^{-1}\Bigl({R_{s,\gamma}(q)-R_{c,0}(q')\over 1+\gamma}\Bigr)
\nonumber \\
& + & {1\over N_a}\sum_{\gamma' =0}\sum_{q'}N_{s,\gamma'}(q')
\Theta_{1+\gamma, 1+\gamma'}\Bigl(R_{s,\gamma}(q)-
R_{s,\gamma'}(q')\Bigr) \, ,
\end{eqnarray}
where

\begin{eqnarray}
\Theta_{\gamma,\gamma'}\Bigl(x\Bigl) & = &
\Theta_{\gamma',\gamma}\Bigl(x\Bigl) =
\delta_{\gamma ,\gamma'}\{2\tan^{-1}\Bigl({x\over 2\gamma}\Bigl)
+ \sum_{l=1}^{\gamma -1}4\tan^{-1}\Bigl({x\over 2l}\Bigl)\}
\nonumber \\
& + & (1-\delta_{\gamma ,\gamma'})\{
2\tan^{-1}\Bigl({x\over |\gamma-\gamma'|}\Bigl)
+ 2\tan^{-1}\Bigl({x\over \gamma+\gamma'}\Bigl)\nonumber \\
& + & \sum_{l=1}^{{\gamma+\gamma'-|\gamma-\gamma'|\over 2}
-1}4\tan^{-1}\Bigl({x\over |\gamma-\gamma'|+2l}\Bigl)\} \, .
\end{eqnarray}
The limits of the pseudo-Brillouin zones, 
$q_{\alpha,\gamma}^{(\pm)}$, associated with the pseudomomentum 
summations are given in Eqs. $(31)-(34)$. 
Although the integral Eqs. $(66)-(68)$ are coupled note
that each of these equations refers to one $\alpha,\gamma$
channel and defines the corresponding $\alpha,\gamma $ rapidity
in terms of other rapidities. 

The key point is that since the eigenvalue $N_{\alpha,\gamma}(q)$
(here $\gamma =0,1,2,3,...$) is common to the whole tower
of $2S^{\alpha }+1$ Hamiltonian eigenstates constructed
from the same $S^{\alpha }$ LWS, also the associate
rapidity eigenvalue $R_{\alpha,\gamma}(q)$ of Eq. $(63)$
is common to these $2S^{\alpha }+1$ Hamiltonian eigenstates.
This follows from the fact that rapidity solutions
$R_{\alpha,\gamma}(q)$ of Eqs. $(66)-(68)$ are functionals
of the $N_{\alpha,\gamma}(q)$ distributions and $\beta $
independent. We thus conclude that the LWS rapidities extracted from
the BA provide full information on the non-LWS's
rapidity operators.

The rapidity eigenvalues of Eqs. $(63)$ and $(65)$ are independent of the 
pseudohole numbers $N^h_{\alpha,\beta}$ and only involve the 
pseudoparticle distributions $N_{\alpha,\gamma}(q)$. The
same holds true for the associate rapidity, pseudohole,
and pseudoparticle operators. Since the $SO(4)$ Hamiltonian
$(2)$ and other physical quantities turn out to be exclusive functionals
of the rapidity operators, in studies involving such quantities
it is often more convenient to use the $\alpha,\gamma$ pseudoparticle 
representation (with $\gamma =0,1,2,3,...$) than the $\alpha,\beta$
pseudohole and $\alpha,\gamma$ heavy-pseudoparticle representation
(with $\gamma =1,2,3,...$). We remind that the set of all
$N^h_{\alpha,\beta}$ and $N_{\alpha,\gamma}$ (with $\gamma =1,2,3,...$)
numbers of a given Hamiltonian eigenstate $(23)$ have the same 
information as the corresponding set of all $N^z_{\alpha }$ and 
$N_{\alpha,\gamma}$ (with $\gamma =0,1,2,3,...$) numbers which
characterize the same state.

By direct insertion in each of Eqs. $(66)-(68)$ of the corresponding 
$q_{\alpha,\gamma}^{(\pm)}$ pseudomomenta we find that the functions 
$R_{\alpha,\gamma}(q)$ have {\it for all} eigenstates the following 
boundary values at the limits of the pseudo-Brillouin zones

\begin{equation}
K(q^{(\pm )}_{c,0}) = \pm\pi \, , \hspace{1cm}
R_{c,0}(q^{(\pm )}_{c,0}) = 0 \, ,
\end{equation}
for $\alpha =c$ and $\gamma =0$ and

\begin{equation}
R_{\alpha,\gamma}(q^{(\pm )}_{\alpha,\gamma}) = \pm\infty \, ,
\end{equation}
for the remaining choices of the quantum numbers $\alpha,\gamma$
other than $c,0$.

It is useful to consider the rapidity functions
$K^{(0)}(q)$ and $R_{\alpha,\gamma}^{(0)}(q)$ which
are the GGS eigenvalues (or GS eigenvalues -- a GS is
a particular case of a GGS) such that

\begin{equation}
\hat{K}(q)|GGS\rangle =
K^{(0)}(q)|GGS\rangle \, ,
\end{equation}
and

\begin{equation}
\hat{R}_{\alpha,\gamma}(q)|GGS\rangle =
R^{(0)}_{\alpha,\gamma}(q)|GGS\rangle \, .
\end{equation}
These are defined by Eqs. $(66)-(68)$ with the pseudomomentum
distribution given by its GGS value which we denote
by $N^{(0)}_{\alpha,\gamma}(q)$ and is such that 

\begin{equation}
\hat{N}_{\alpha,\gamma}(q)|GGS\rangle =
N^{(0)}_{\alpha,\gamma}(q)|GGS\rangle \, .
\end{equation}

Following the GGS expression $(40)$, $N^{(0)}_{\alpha,\gamma}(q)$
has the simple free Fermi-like form

\begin{eqnarray}
N^{(0)}_{\alpha,\gamma}(q) & = & \Theta\Bigl(
C_{\alpha,\gamma}[q_{F\alpha,\gamma}^{(+)}
- q]\Bigl) \, , \hspace{1cm} 0<q<q_{\alpha,\gamma}^{(+)}
\nonumber \\
& = & \Theta\Bigl(C_{\alpha,\gamma}[q - 
q_{F\alpha,\gamma}^{(-)}]\Bigl) \, , 
\hspace{1cm} q_{\alpha,\gamma}^{(-)}<q<0 \, .
\end{eqnarray}
For the particular case of a GS these distributions read

\begin{eqnarray}
N^{(0)}_{\alpha,0}(q) & = & \Theta\Bigl(q_{F\alpha,0}^{(+)}
- q \Bigl) \, , \hspace{1cm} 0<q<q_{\alpha,0}^{(+)}
\nonumber \\
& = & \Theta\Bigl(q - q_{F\alpha,0}^{(-)}\Bigl) \, , 
\hspace{1cm} q_{\alpha,0}^{(-)}<q<0 \, ,
\nonumber \\
N^{(0)}_{\alpha,\gamma}(q) & = & 0 \, , \hspace{2cm} \gamma > 0  \, .
\end{eqnarray}

By inserting in Eqs. $(66)-(68)$ the GGS distribution $(75)$ 
we define the GGS rapidities $K^{(0)}(q)$ and 
$R_{\alpha,\gamma}^{(0)}(q)$ in terms of their inverse functions 
with the result

\begin{equation}
q = \int_{0}^{K^{(0)}(q)}
dk 2\pi\rho_{c,0}(k) \, , 
\end{equation}

\begin{equation}
q = \int_{0}^{R^{(0)}_{\alpha,\gamma}(q)}
dr 2\pi\rho_{\alpha,\gamma}(r) \, , 
\end{equation}
where the fuctions $2\pi\rho_{c,0}(k)$ and $2\pi\rho_{\alpha,\gamma}(r)$
are the solutions of the integral equations (B4)-(B6) of Appendix B.
In that Appendix we study the rapidity solutions of Eqs.
$(66)-(68)$ both for GGS's and for Hamiltonian eigenstates differing
from a GGS by a small density of excited pseudoparticles.
In equations (B4)-(B6) the parameters $Q^{(\pm)}$ and 
$r^{(\pm)}_{\alpha,\gamma }$ are defined combining the
following equations

\begin{equation}
Q^{(\pm)} = K^{(0)}(q^{(\pm )}_{Fc,0}) \, , \hspace{1cm}
r^{(\pm)}_{\alpha,\gamma} = R^{(0)}_{\alpha,\gamma}(q^{(\pm 
)}_{F\alpha,\gamma}) \, ,
\end{equation}
with equations $(77)$ and $(78)$. We also introduce the associate 
parameters

\begin{equation}
Q = K^{(0)}(q_{Fc,0}) \, , \hspace{1cm}
r_{\alpha,\gamma} = R^{(0)}_{\alpha,\gamma}(q_{F\alpha,\gamma}) \, .
\end{equation}

The system of coupled algebraic equations $(66)-(68)$ is obtained 
from the BA solution (see Ref. \cite{Takahashi} and Appendix A 
which connects our rapidity representation to the notation used in that
reference), yet also describes states out of that solution. Given the 
configuration of pseudoparticle quantum numbers that describes 
each Hamiltonian eigenstate of form $(23)$, these equations fully 
define the corresponding rapidities. In the thermodynamic limit, we 
can take the pseudomomentum continuum limit $q_j\rightarrow q$ and 
the real part of the rapidities become functions of $q$ which we have 
called here $R_{\alpha,\gamma} (q)$. In that limit the BA system of 
algebraic equations are replaced by the system of infinite coupled 
integral equations presented in Appendix B. 

We have mentioned that the BA solution is most naturally expressed 
in the pseudoparticle basis. One reflection of this is the simple 
expression for the Hamiltonian $(1)$ in that basis. The BA energy 
expression given in Refs. $\cite{Essler,Takahashi}$ for the LWS's 
of the $\eta$-spin and spin algebras, which we denote by $E_{BA}$,
can be generalized to all Hamiltonian eigenstates. For that
we replace the $-S^{\alpha}_z$ dependence of $E_{BA}$ by a
$S^{\alpha}$

\begin{equation}
E\equiv E_{BA}(-S^{\alpha}_z\rightarrow S^{\alpha}) \, .
\end{equation}
Combining Eq. $(19)$ with the energy expression $(81)$
leads then to the following expression for the Hamiltonian $(1)$

\begin{equation}   
\hat{H} = \hat{H}_{SO(4)} + 2\mu\sum_{\beta}\beta\hat{N}^h_{c,\beta} 
+ 2\mu_0 H\sum_{\beta}\beta\hat{N}^h_{s,\beta} \, ,
\end{equation}
where

\begin{eqnarray}
\hat{H}_{SO(4)} & = & -2t\sum_{q}\hat{N}_{c,0}(q)\cos [\hat{K}(q)]
+ 2t\sum_{q,\gamma =1}\hat{N}_{c,\gamma}(q)
\sum_{j=\pm 1}\sqrt{1-[u(\hat{R}_{c,\gamma}(q)-ji\gamma)]^2}
\nonumber \\
& + & U[{N_a\over 4} - {\hat{N}_{c,0}\over 2}
- \sum_{\gamma =1}\gamma \hat{N}_{c,\gamma}] \, .
\end{eqnarray}
This is the rigurous pseudoparticle expression of the Hamiltonian 
$(2)$ at all energy scales. It gives the exact expression 
of that Hamiltonian in the full Hilbert space.
Despite its simple appearance, the Hamiltonian $(83)$ 
describes a many-pseudoparticle problem. The reason is that the 
expression of the rapidity operator in terms of 
the operators $\hat{N}_{\alpha,\gamma}(q)$ contains many-pseudoparticle
interacting terms.

To specify the Hamiltonian $(83)$ completely, we must indicate
the equations that define the rapidity operator eigenvalues
in terms of the pseudoparticle momentum distribution operators.
In the case of the present model these are Eqs. $(66)-(68)$.

Unsurprisingly, it is difficult to solve the BA operator equations 
$(66)-(68)$ directly and to obtain the explicit expression for the rapidity 
operators in terms of the pseudomomentum distribution operators 
$(15)$. In contrast, it is easier to calculate their normal-ordered 
expression in terms of the normal-ordered operators. 
The rapidity operators $\hat{R}_{\alpha,\gamma}(q)$
contain {\it all} information about the many-pseudoparticle
interactions of the quantum-liquid Hamiltonian. There are two 
fundamental properties which imply the central role that the  
rapidity operators of Eqs. $(72)$ and $(73)$ have in the present 
quantum problem:

(a) As we find in Sec. V, each of the normal-ordered rapidity 
operator $:\hat{R}_{\alpha}(q):$ (relative to the suitable
GGS or GS) can be written exclusively in terms of the 
pseudomomentum distribution operators $(15)$;

(b) The normal-ordered version of the $SO(4)$ Hamiltonian $(83)$ can 
be written, exclusively, in terms of the pseudomomentum 
distribution operators $(15)$, but all the corresponding 
many-pseudoparticle interaction terms can be written in terms 
of the rapidity operators $:\hat{R}_{\alpha,\gamma}(q): $.
It follows that the rapidity operators commute with the Hamiltonian. 

In the ensuing section, we introduce the pseudoparticle perturbation 
theory which leads to the normal-ordered expressions for the 
Hamiltonian and rapidity operators. Despite the non-interacting 
form of the Hamiltonian eigenstates $(23)$, the normal-ordered 
Hamiltonian includes pseudoparticle interaction terms and is, therefore, 
a many-pseudoparticle operator. However, we find in Sec. V that
these pseudoparticle interactions have a pure forward-scattering, 
zero-momentum transfer, character. 

\section{PSEUDOPARTICLE HAMILTONIAN AND PERTURBATION THEORY}

We will be mostly interested in GS - GGS transitions 
followed by pseudoparticle - pseudohole excitations
involving a small density of $\alpha,\gamma$ pseudoparticles 
relative to both the GS and GGS distributions.  Let us then 
introduce the normal-ordered pseudomomentum distribution
operator

\begin{equation}
:\hat{N}_{\alpha,\gamma}(q): =
\hat{N}_{\alpha,\gamma}(q) - N^{(0)}_{\alpha,\gamma}(q) \, .
\end{equation}
We emphasize that when we choose $N^{(0)}_{\alpha,\gamma}(q)$ to be 
the GS pseudomomentum distribution $(76)$ we have that 

\begin{equation}
:\hat{N}_{\alpha,\gamma}(q): =
\hat{N}_{\alpha,\gamma}(q) \, , \hspace{2cm} \gamma > 0 \, .
\end{equation}

The normal-ordered distribution $(84)$ obeys the eigenvalue equation

\begin{equation}
:\hat{N}_{\alpha,\gamma}(q):|\psi\rangle =
\delta N_{\alpha,\gamma}(q)|\psi\rangle \, .
\end{equation}
We also consider the normal-ordered rapidity operator

\begin{equation}
:\hat{R}_{\alpha,\gamma}(q): =
\hat{R}_{\alpha,\gamma}(q) - R^{(0)}_{\alpha,\gamma}(q) \, ,
\end{equation}
where $R_{\alpha,\gamma}^0(q)$ is the GGS eigenvalue
of Eq. $(73)$.

In the pseudoparticle basis the normal-ordered rapidity 
operators $:\hat{R}_{\alpha,\gamma}(q):$ contain
an infinite number of terms, as we shall demonstrate below. The first of
these terms is linear in the pseudomomentum distribution
operator $:\hat{N}_{\alpha,\gamma}(q):$, $(84)$, whereas the remaining terms 
consist of products of two, three,....., until infinity, 
of these operators. The number of $:\hat{N}_{\alpha,\gamma}(q):$ 
operators which appears in these products equals the order of the
scattering in the corresponding rapidity term.

A remarkable property is that in the pseudoparticle 
basis the seemingly ``non-perturbative'' quantum liquids 
become {\it perturbative}: while the two-electron
forward scattering amplitudes and vertices diverge, the
two-pseudoparticle $f$ functions [given by Eq. $(117)$ 
below] are finite. 

By ``perturbative'' we also mean here the 
following: at each point of parameter space (canonical
ensemble) the excited eigenstates are of form $(23)$ 
and correspond to quantum-number configurations involving a 
density of excited pseudoholes and heavy pseudoparticles relative to the GS 
configuration $(46)$ and $(76)$. We introduce the following
density 

\begin{equation}
n_{ex}' = \sum_{\alpha,\beta}n_{ex}^{\alpha,\beta} +
\sum_{\alpha,\gamma =1}n_{ex}^{\alpha,\gamma} \, ,
\end{equation}
which is kept small. Here

\begin{equation}
n_{ex}^{\alpha,\beta} = {1\over {N_a}}\sum_{q}
[1 - N_{\alpha,\beta}^{h,(0)}(q)]\delta N_{\alpha, \beta}(q) \, , 
\end{equation}
and

\begin{equation}
n_{ex}^{\alpha,\gamma} = {1\over {N_a}}\sum_{q}
N_{\alpha,\gamma}(q) \, ,  
\end{equation}
define the densities of excited $\alpha,\beta$ pseudoholes
and $\alpha,\gamma$ heavy pseudoparticles, respectively, 
associated with the Hamiltonian eigenstate $|\psi\rangle$.

In Eq. $(89)$ $N_{\alpha,\beta}^{h,(0)}(q)$ is the GS pseudohole
distribution which reads

\begin{eqnarray}
N_{\alpha,-{1\over 2}}^{h,(0)}(q) & = & 1 - N_{\alpha,0}^{(0)}(q)
\, ,\nonumber \\
N_{\alpha,+{1\over 2}}^{h,(0)}(q) & = & 0 \, ,
\end{eqnarray}
if the GS is a $S^{\alpha }$ LWS and 

\begin{eqnarray}
N_{\alpha,-{1\over 2}}^{h,(0)}(q) & = & 0
\, ,\nonumber \\
N_{\alpha,+{1\over 2}}^{h,(0)}(q) & = & 1 - N_{\alpha,0}^{(0)}(q) \, ,
\end{eqnarray}
if the GS is a $S^{\alpha }$ HWS, where $N_{\alpha,0}^{(0)}(q)$
is the $\alpha,0$ pseudoparticle GS distribution $(76)$.
Our perturbation theory refers to general GS transitions 
to Hamiltonian eigenstates $|\psi\rangle$ involving
(a) a GS - GGS transition which costs the energy $\omega_0$ [given
by Eq. $(124)$ below] and (b) a Landau-liquid 
pseudoparticle-pseudohole excitation around the GGS. The pseudoparticle 
perturbation theory corresponds to such GS transitions
with excitation energy $\omega $ such that the energy
$(\omega -\omega_0)$ is small. Since the initial state is a GS we have 
assumed the initial GS distribution $(76)$ for the $\alpha,\gamma$ heavy 
pseudoparticles. Following Eq. $(76)$ we thus have that $N_{\alpha,\gamma}(q) =
\delta N_{\alpha,\gamma}(q)$ in the rhs of Eq. $(90)$.  

The expectation values of the $SO(4)$ Hamiltonian 
$(83)$ in the final states $|\psi\rangle$ are functions of the density 
of excited pseudoparticles only. This density is given by

\begin{equation}
n_{ex} = \sum_{\alpha,\gamma =0}n_{ex}^{\alpha,\gamma} \, ,
\end{equation}
where $n_{ex}^{\alpha,\gamma}$ is given by

\begin{equation}
n_{ex}^{\alpha,0} = {1\over {N_a}}\sum_{q}
[1 - N_{\alpha,0}^{(0)}(q)]\delta N_{\alpha,0}(q) \, , 
\end{equation}
for $\gamma =0$ and by $(90)$ otherwise. When both the pseudoparticle 
density $(93)$ and $n^z={\sum_{\alpha}N^z_{\alpha }\over N_a}$ [see
Eq. $(13)$] are small implies that the density $(88)$ is also small.
On the other hand, the pseudoparticle density $(93)$ 
is small provided the density $(88)$ is also small. 
This then implies that all the densities $n_{ex}^{\alpha,\gamma}$ are 
small and that we can expand the expectation values in these densities. 
The perturbative character of the quantum liquid rests on the 
fact that the evaluation of the expectation values up to the
$j^{th}$ order in the densities $(93)$ requires considering only the 
corresponding operator terms of scattering orders less than or
equal to $j$. This follows from the linearity of the density of 
excited $\alpha,\gamma$ pseudoparticles, which are the elementary 
``particles'' of the quantum liquid, in 

\begin{equation}
\delta N_{\alpha,\gamma}(q)=\langle\psi|:\hat{N}_{\alpha,\gamma}(q):
|\psi\rangle \, ,
\end{equation}
and from the form of $(90)$ and $(94)$. The perturbative character 
of the quantum liquid implies, for example, that, to second order 
in the density of excited pseudoparticles, the energy involves only one-
and two-pseudoparticle Hamiltonian terms, 
as in the case of the quasiparticle terms of a Fermi-liquid 
energy functional \cite{Pines,Baym}. In this case the corresponding
energy $(\omega-\omega_0)$ is small.

It follows from the eigenfunction Eq. $(86)$ that the problem 
of using the rapidity Eqs. $(66)-(68)$ to derive the 
expression of the operators $:\hat{R}_{\alpha,\gamma}(q):$ 
in terms of the operators $:\hat{N}_{\alpha,\gamma}(q):$ is equivalent 
to the problem of evaluating the corresponding expansion
of the rapidity eigenvalues $\delta R_{\alpha,\gamma}(q)$ in 
terms of the pseudoparticle deviations $\delta N_{\alpha,\gamma}(q)$ 
$(95)$. This last problem, which leads to the Landau-liquid expansions, 
was studied in Refs. \cite{Carmelo92b,Carmelo92c,Carmelo94b} for 
the case of the Hilbert subspace spanned by the states I. 
Furthermore, we emphasize that it is the perturbative character of 
the pseudoparticle basis which {\it justifies} the validity of 
these Landau-liquid deviation expansions 
\cite{Carmelo96,Carmelo91,Carmelo92b,Carmelo92c,Carmelo94b}.

Since a GS is a particular case of a GGS, in the following we consider the 
general case of normal ordering relative to a GGS. Only by using as
reference state a GGS (with heavy-pseudoparticle occupancy) and obtaining 
the GS expressions as a particular limit of the GGS general expressions we 
can be shure to arrive to the correct GS results. Later on 
we will specify when the ordering refers to a GS.

We can derive the expressions for the rapidity operators 
$:{\hat{R}}_{\alpha, \gamma}(q):$. This corresponds to expanding the
expressions of the operators $:{\hat{R}}_{\alpha, \gamma}(q):$ in terms
of increasing pseudoparticle scattering order. It is convenient 
to define these expressions through the operators 
$:{\hat{X}}_{\alpha, \gamma}(q):$. These are related to the rapidity 
operators as follows

\begin{equation}
:{\hat{R}}_{\alpha, \gamma}(q): = 
R_{\alpha, \gamma}^0(q + :{\hat{X}}_{\alpha, \gamma}(q):) 
- R_{\alpha, \gamma}^0(q) \, ,
\end{equation}
where $R_{\alpha,\gamma}^{(0)}(q)$ is the suitable  
GGS eigenvalue of ${\hat{R}}_{\alpha,\gamma}(q)$
[see Eq. $(73)$].

The operators $:\hat{X}_{\alpha,\gamma}(q):$ contain the same
information as the rapidity operators, and involve, exclusively, the 
two-pseudoparticle phase shifts $\Phi_{\alpha,\gamma;\alpha '
,\gamma'}(q,q ')$ we define below. In Appendix B we introduce
$(96)$ in the BA equations $(66)-(68)$ and expand in the scattering order. 
This leads to

\begin{equation}
:\hat{X}_{\alpha,\gamma}(q): = \sum_{j=1}^{\infty}
\hat{X}_{\alpha,\gamma}^{(j)}(q) \, ,
\end{equation}
where $j$ gives the scattering order of the operator
term $\hat{X}_{\alpha,\gamma}^{(j)}(q)$. For the
first-order term we find

\begin{equation}
\hat{X}_{\alpha,\gamma}^{(1)}(q) = {2\pi\over {N_a}}
\sum_{q',\alpha',\gamma'}
\Phi_{\alpha,\gamma;\alpha ',\gamma'}(q,q ')
:\hat{N}_{\alpha',\gamma'}(q'):
\, ,
\end{equation}
where the phase-shift expressions are given below.
We emphasize that while the expressions for the phase shifts 
$\Phi_{\alpha,\gamma;\alpha ',\gamma'}(q,q ')$ are specific to each model
because they involve the spectral parameters, the {\it form} of 
the operator term $\hat{X}_{\alpha,\gamma}^{(1)}(q)$ $(98)$ is 
{\it universal} and refers to all the solvable electronic multicomponent 
quantum liquids. 

It is useful to introduce the phase shifts 
$\tilde{\Phi}_{\alpha,\gamma;\alpha ',\gamma'}$ such that

\begin{equation}
\tilde{\Phi}_{c,0;c,0}(k,k') = 
\bar{\Phi }_{c,0;c,0}\left({\sin k\over u},
{\sin k'\over u}\right) \, ,
\end{equation}

\begin{equation}
\tilde{\Phi}_{c,0;\alpha ,\gamma}(k,r') = 
\bar{\Phi }_{c,0;\alpha ,\gamma}\left({\sin k\over u},
r'\right) \, ,
\end{equation}

\begin{equation}
\tilde{\Phi}_{\alpha ,\gamma;c,0}(r,k') = 
\bar{\Phi }_{\alpha ,\gamma;c,0}\left(r,
{\sin k'\over u}\right) \, ,
\end{equation}

\begin{equation}
\tilde{\Phi}_{\alpha,\gamma;\alpha,\gamma'}\left(r,r'\right) = 
\bar{\Phi }_{\alpha,\gamma;\alpha,\gamma'}\left(r,r'\right) \, ,
\end{equation}
where $\gamma >0$ and the phase shifts 
$\bar{\Phi }_{\alpha,\gamma;\alpha ',\gamma'}$ 
are defined by the integral equations (B30)-(B40) of Appendix B.

The two-pseudparticle phase shifts can be defined in terms of
the phase shifts $\bar{\Phi }_{\alpha,\gamma;\alpha',\gamma'}$ as 
follows

\begin{equation}
\Phi_{\alpha,\gamma;\alpha',\gamma'}(q,q') = 
\bar{\Phi }_{\alpha,\gamma;\alpha',\gamma'}
\left(R^{(0)}_{\alpha,\gamma}(q),
R^{(0)}_{\alpha,\gamma}(q')\right) \, .
\end{equation}
The quantity $\Phi_{\alpha,\gamma;\alpha',\gamma'}(q,q')$ 
represents the shift in the phase of the $\alpha ',\gamma '$ 
pseudoparticle of pseudomomentum $q'$ due to a zero-momentum 
forward-scattering collision with the $\alpha ,\gamma$ pseudoparticle
of pseudomomentum $q$.

In Appendix C we use the Hamiltonian expression $(83)$ in terms of 
the rapidity operators to derive the expression 
for the normal-ordered Hamiltonian. We find that
in normal order relative to the suitable GGS 
(or GS) the Hamiltonian ${\hat{H}}_{SO(4)}$, Eq. $(83)$, can 
be written as

\begin{equation}
:{\hat{H}}_{SO(4)}: = \sum_{j=1}^{\infty}\hat{H}^{(j)} \, . 
\end{equation}
For example, the first and second pseudoparticle-scattering order 
terms read
 
\begin{equation}
\hat{H}^{(1)}=\sum_{q,\alpha,\gamma}
\epsilon^0_{\alpha ,\gamma}(q):\hat{N}_{\alpha ,\gamma}(q): \, ,
\end{equation}
and 

\begin{equation}
\hat{H}^{(2)}={1\over {N_a}}\sum_{q,\alpha ,\gamma} 
\sum_{q',\alpha',\gamma'}{1\over 2}f_{\alpha,\gamma;
\alpha',\gamma'}(q,q') :\hat{N}_{\alpha ,\gamma}(q):
:\hat{N}_{\alpha' ,\gamma'}(q'): \, , 
\end{equation}
respectively. All the remaining higher-order operator terms of 
expressions $(97)$ and $(104)$ can be obtained by combining 
the rapidity equations $(66)-(68)$ and the Hamiltonian expression 
$(83)$.

In Appendix C it is shown that the pseudoparticle bands 
$\epsilon^0_{\alpha ,\gamma}(q)$ can either be expressed
in terms of the phase shifts $(103)$, with the result

\begin{equation}
\epsilon_{c,0}^0(q) = -{U\over 2} -2t\cos K^{(0)}(q) +
2t\int_{Q^{(-)}}^{Q^{(+)}}dk\widetilde{\Phi }_{c,0;c,0}
\left(k,K^{(0)}(q)\right)\sin k \, ,
\end{equation}

\begin{equation}
\epsilon_{c,\gamma}^0(q) = -\gamma U 
+ 4t Re \sqrt{1 - u^2[R^{(0)}_{c,\gamma}(q) - i\gamma]^2}
+ 2t\int_{Q^{(-)}}^{Q^{(+)}}dk\widetilde{\Phi }_{c,0;c,\gamma}
\left(k,R^{(0)}_{c,\gamma}(q)\right)\sin k \, ,
\end{equation}
for $\gamma >0$ and

\begin{equation}
\epsilon_{s,\gamma}^0(q) = 2t\int_{Q^{(-)}}^{Q^{(+)}}dk
\widetilde{\Phi }_{c,0;s,\gamma}
\left(k,R^{(0)}_{s,\gamma}(q)\right)\sin k \, ,
\end{equation}
for $\gamma =0,1,2,...$, or simply as

\begin{equation}
\epsilon_{c,0}^0(q) = \epsilon_{c,0}^0(\pi) + \int_{\pi}^{K^{(0)}(q)}
dk 2t\eta_{c,0}(k) \, , 
\end{equation}

\begin{equation}
\epsilon_{c,\gamma}^0(q) =
\int_{\infty}^{R^{(0)}_{c,\gamma}(q)}
dr 2t\eta_{c,\gamma}(r) \, , 
\end{equation}
for $\gamma >0$ and

\begin{equation}
\epsilon_{s,\gamma}^0(q) =
\int_{\infty}^{R^{(0)}_{s,\gamma}(q)}
dr 2t\eta_{s,\gamma}(r) \, , 
\end{equation}
for $\gamma =0,1,2,...$, where the fuctions $2t\eta_{c,0}(k)$ and 
$2t\eta_{\alpha,\gamma}(r)$
are the solutions of the integral equations (C7)-(C9) of
Appendix C. The GGS (or GS) rapidity functions $K^{(0)}(q)$ and 
$R^{(0)}_{\alpha,\gamma}(q)$ are defined by Eqs. $(77)$, $(78)$
combined with Eqs. (B4)-(B6) of Appendix B. Combining Eqs. $(70)$ and 
$(71)$ with the band expressions we find that the bands $(108)$, 
$(111)$ and $(109)$, $(112)$ vanish at the limits of the
pseudo-Brillouin zones, i.e.

\begin{equation}
\epsilon_{\alpha,\gamma}^0(q^{(\pm)}_{\alpha,\gamma}) = 0 \, . 
\end{equation}

The associate group velocities, $v_{\alpha ,\gamma}(q)$,
are given by

\begin{equation}
v_{\alpha ,\gamma}(q) = 
{d\epsilon^0_{\alpha ,\gamma}(q) \over {dq}} \, . 
\end{equation}

The two-pseudoparticle forward-scattering phase shifts 
$\Phi_{\alpha,\gamma;\alpha',\gamma'}(q,q')$ defined by 
Eq. $(103)$ and velocities 

\begin{equation}
v_{\alpha ,\gamma}\equiv v_{\alpha ,\gamma}(q_{F\alpha ,\gamma}) \, , 
\end{equation}
play an important role in the physical quantities when criticality
is appoached \cite{Carmelo96b,Carmelo96d}.

All $\hat{X}_{\alpha,\gamma}^{(j)}(q)$ terms of the rhs of Eq. $(97)$
are such that both the $f$ functions of the rhs of Eq. $(106)$ 
and all the remaining higher order coefficients associated with the 
operators $\hat{H}^{(j)}$ of order $j>1$ have {\it universal} 
forms in terms of the {\it two-pseudoparticle} phase shifts and 
pseudomomentum derivatives of the bands and coefficients 
of order $<j$. This follows from the fact that the $S$-matrix for 
$j$-pseudoparticle scattering factorizes into two-pseudoparticle 
scattering matrices, as in the case of the usual BA
$S$-matrix \cite{Essler,Korepin79,Zamo,Andrei}. For
example, we show in Appendix C that although the second-order term 
$\hat{H}^{(2)}$ of Eq. $(106)$ involves an integral over the second-order
function $\hat{X}_{\alpha,\gamma}^{(2)}(q)$ [see Eq. (C14) of
Appendix C], this function is such that $\hat{H}^{(2)}$
can be written {\it exclusively} in terms of the
first-order functions $(98)$,
  
\begin{equation}
\hat{H}^{(2)} = \sum_{q,\alpha,\gamma}v_{\alpha,\gamma} (q)
\hat{X}_{\alpha,\gamma}^{(1)}(q) :\hat{N}_{\alpha,\gamma}(q):
+ {N_a\over {2\pi}}\sum_{\alpha,\gamma}\theta (N_{\alpha,\gamma})
{|v_{\alpha,\gamma}|\over 2}
\sum_{j=\pm 1}[\hat{X}_{\alpha,\gamma}^{(1)}(jq_{F\alpha,\gamma})]^2 \, ,
\end{equation} 
where $|v_{\alpha,\gamma}|=C_{\alpha,\gamma}v_{\alpha,\gamma}$,
$\theta (x)=1$ for $x>0$, and $\theta (x)=0$ for $x\leq 0$
and the ``Landau'' $f$ functions, $f_{\alpha,\gamma;\gamma'
,\alpha'}(q,q')$, are found in that Appendix to have universal 
form which reads

\begin{eqnarray}
f_{\alpha,\gamma;\alpha',\gamma'}(q,q') & = & 2\pi 
v_{\alpha,\gamma}(q)\Phi_{\alpha,\gamma;\alpha',\gamma'}(q,q')+
2\pi v_{\alpha',\gamma'}(q')
\Phi_{\alpha',\gamma';\alpha,\gamma}(q',q) \nonumber \\
& + & 2\pi\sum_{j=\pm 1} 
\sum_{\alpha''}\sum_{\gamma''=0}^{\infty}
\theta (N_{\alpha'',\gamma''})C_{\alpha'',\gamma''}v_{\alpha'',\gamma''}
\Phi_{\alpha'',\gamma'';\alpha,\gamma}(jq_{F\alpha''
,\gamma''},q)\Phi_{\alpha'',\gamma'';\alpha',\gamma'}
(jq_{F\alpha'',\gamma''},q') \, ,
\end{eqnarray}
where the pseudoparticle group velocities are given by 
Eqs. $(114)$ and $(115)$. 

We can also write the expression of $\hat{H}^{(j)}$ for higher 
scattering orders $j>2$. The main feature is that for all
sub-canonical ensembles, energy scales, and pseudoparticle 
scattering orders, the Landau-liquid terms of that Hamiltonian 
involve only zero-momentum forward-scattering.

Let us consider that the normal-ordering of the
Hamiltonian $(104)$ refers to the initial GS. In this case, and 
in contrast to the low-energy Landau theory studied in Refs. 
\cite{Carmelo96,Carmelo91,Carmelo92b,Carmelo92c,Carmelo94b},
the Hamiltonian $(104)$ describes finite-energy transitions
involving a small density of excited pseudoparticles.
Furthermore, a finite-size analysis \cite{Carmelo96d}
confirms that the excitations which control the response and transport 
properties at finite energies involve a small density of excited 
pseudoparticles. Fortunately, their description is thus within the 
range of our generalized pseudoparticle perturbation theory.
It is convenient to define the normal ordering relative to the initial GS 
but with energy shifted by

\begin{equation}
\omega_0 = \lim_{N_a\to\infty}[E_{GGS}-E_{GS}] \, . 
\end{equation}
This leads to a small $(\omega-\omega_0)$ energy description.
Fortunately, the relevant GS - GGS transitions correspond to energy 
values $\omega_0$ associated with small-pseudoparticle-density 
excitations \cite{Carmelo96b,Carmelo96d}. For each of these $\omega_0$ 
values there is a low-energy $(\omega -\omega_0)$ pseudoparticle-perturbation
theory. The associate Hilbert subspace is spanned by the set of GGS's
of energy $\omega_0$ and by states originated from these 
by pseudoparticle-pseudohole processes. The latter states
have small momentum and low $(\omega -\omega_0)$ energy relatively
to the initial GS (we have shifted its energy by $\omega_0$).
Normal-ordering relative to the energy-$\omega_0$ shifted initial GS 
is equivalent to separate the problem $(104)$ into (a) the GS - GGS 
transition of finite-energy $(118)$ and (b) a low $(\omega -\omega_0)$ energy 
theory associated with both the corrections of order $1/N_a$ to the 
GS - GGS-transition energy $(118)$ and the Landau-liquid pseudoparticle-pseudohole 
processes around the final GGS. In most cases the excitation energy
$(118)$ associated with the GS - GGS transitions is finite. 
The finite energies are the energy gaps of the states 
II and (or) non-LWS's and non-HWS's relative to the 
initial GS.

At finite values of $S^c_z$ and $S^s_z$ the low-energy Hilbert 
space is entirely spanned by states I \cite{Carmelo94,Carmelo96,Carmelo94b}.
We note that for these states fixing the electron numbers fixes 
the pseudohole numbers \cite{Carmelo96} and the $\alpha,\gamma$ 
bands are empty for $\gamma >0$. Therefore, if we fix the electron 
numbers we have that at energy scales smaller 
than the gaps $(118)$ for the states II and non-LWS's and non-HWS's 
only Landau-liquid excitations (b) are allowed.
This justifies the Landau-liquid character of the problem at
low energies \cite{Carmelo92,Carmelo93,Carmelo94,Carmelo96}.
Obviously, if we change the electron numbers, there occur
at low energy GS - GS transitions which are particular
cases of the above GS - GGS transitions (a). Understanding
these GS - GS transitions permits one to express the 
electron in terms of pseudoholes and topological momentum 
shifts \cite{Carmelo96,Carmelo96c}. 

Let us evaluate the general $\omega_0$ expression for GS - GGS 
transitions such that the final sub-canonical ensemble 
is characterized by vanishing or small values of 
$N^h_{\alpha ,{-l_{\alpha}\over 2}}/N^h_{\alpha }$. Here 
$l_{\alpha }=\pm 1$ is defined by Eq. $(7)$. 
We also assume that both the initial GS and final GGS 
correspond to canonical ensembles such that $S^{\alpha}_z\neq 0$
and belonging the same sector of parameter space, i.e. with the
same $l_{\alpha }$ numbers. We note that since GS's are
always $S^{\alpha}$ LWS's I or HWS's I there are neither 
$\alpha ,{-l_{\alpha}\over 2}$ pseudoholes [see Eqs. $(37)$ 
and $(38)$] nor $\alpha,\gamma$ heavy pseudoparticles ($\gamma >0)$ 
in the initial GS and that $N^z_{\alpha }=0$ [see Eq. $(13)$]
for that state. From Eqs. $(104)-(106)$, we find for the GS - GGS gap

\begin{equation}
\omega_0 = \sum_{\alpha } \epsilon^0_{\alpha ,0}(q_{F\alpha,0})
\Delta N_{\alpha ,0} + \sum_{\gamma =1}
\epsilon^0_{s ,\gamma }(0)N_{s ,\gamma} 
+ 2\mu\Delta S^c_z + 2\mu_0 H\Delta S^s_z
\, ,
\end{equation}
where the band expressions at the pseudo-Fermi points
are given in Eq. (C12) of Appendix C. Inserting in Eqs. 
$(11)$ both the pseudohole expressions $(12)$ and 
the expressions $S^c_z=-{1\over 2}[N_a-N]$ and
$S^s_z=-{1\over 2}[N_{\uparrow}-N_{\downarrow}]$ we find
after use of Eqs. $(13)$ and $(20)$ the following
general expressions for $\Delta N_{c,0}$ and $\Delta N_{s,0}$

\begin{equation}
\Delta N_{c,0} = - 2l_c\Delta S^c_z - 2N^z_c - \sum_{\gamma =1}2\gamma 
N_{c,\gamma } = -[N^h_{c ,{-l_{c}\over 2}}
- \Delta N^h_{c ,{l_{c}\over 2}}] - 2N^z_c - \sum_{\gamma =1}2\gamma 
N_{c,\gamma }  \, ,
\end{equation}
and

\begin{eqnarray}
\Delta N_{s,0} & = & - \sum_{\alpha }
[l_{\alpha}\Delta S^{\alpha}_z + N^z_{\alpha}]
- \sum_{\gamma =1}\gamma N_{c,\gamma }  
- \sum_{\gamma =1}(1+\gamma) N_{s,\gamma } \nonumber \\
& = & - {1\over 2}\sum_{\alpha }[N^h_{\alpha ,{-l_{\alpha}\over 2}}
- \Delta N^h_{\alpha ,{l_{\alpha}\over 2}} + 2N^z_{\alpha}] 
- \sum_{\gamma =1}\gamma N_{c,\gamma } 
- \sum_{\gamma =1}(1+\gamma) N_{s,\gamma } \, .
\end{eqnarray}
Generalization of the chemical-potential and 
magnetic-field expressions $(47)$ of Ref. \cite{Carmelo92b} to 
all sectores of parameter space leads to

\begin{equation}
|\mu | = - \epsilon^0_{c,0}(q_{Fc,0})
- {\epsilon^0_{s,0}(q_{Fs,0})\over 2}  
\, , \hspace{1cm}
2\mu_0 |H| = - \epsilon^0_{s,0}(q_{Fs,0}) \, .
\end{equation}

From the use of Eqs. $(20)$ and $(120)-(122)$ in Eq. $(119)$ 
leads to the following gap expression 

\begin{equation}
\omega_0 = 2|\mu|N^h_{c,-{l_c\over 2}} + 2\mu_0|H|[N^h_{s,-{l_s\over 2}}
+ \sum_{\gamma =1}^{\infty}N_{s,\gamma}] 
+ \sum_{\gamma =1}\epsilon^0_{s ,\gamma }(0)
N_{s ,\gamma} \, ,
\end{equation}
which is the general expression for the gap for GS - GGS
transitions. It follows from Eqs. $(13)$, $(19)$, and $(20)$ that the
gap expression $(123)$ can be rewritten as

\begin{equation}
\omega_0 = 2|\mu|\Bigl(\sum_{\gamma =1}\gamma  
N_{c,\gamma} + N^z_c\Bigl)
+ 2\mu_0|H|\Bigl(\sum_{\gamma =1}(1+\gamma ) 
N_{s,\gamma} + N^z_s\Bigl)
+ \sum_{\gamma =1}\epsilon^0_{s ,\gamma }(0)
N_{s ,\gamma} \, .
\end{equation}

In reference \cite{Carmelo96d} the concept of GGS is generalized to 
$\alpha,\gamma >0$ pseudoparticle filled seas with pseudo-Fermi points
either of the form (a) $q_{F\alpha,\gamma}={\pi[d_{\alpha,\gamma}-
N_{\alpha,\gamma}]\over N_a}$ or (b) $q_{F\alpha,\gamma}=
{\pi N_{\alpha,\gamma}\over N_a}$. We note that our definition
$(43)$ corresponds to the case (a) for $\alpha =c$ and (b)
for $\alpha =s$. While the first two terms of the rhs of Eq. $(124)$
are the same for either case, the third term vanishes
for the case (a) [this follows from Eq. $(113)$] and reads 
$\epsilon^0_{\alpha,\gamma }(0)$ for $\alpha,\gamma $ seas 
corresponding to the case (b). This explains the form of the latter 
term in Eq. $(124)$ which refers only to the $s,\gamma $ 
pseudoparticles.

For GS - GGS transitions to pure LWS's or HWS's 
II the use of Eq. $(13)$ leads to

\begin{equation}
N^z_{\alpha } = 0 \, ,
\end{equation}
for both $\alpha =c,s$. It follows that for such transitions
the gap is given by expression $(124)$ with $N^z_c = 0$ and 
$N^z_s = 0$. Therefore, the single ($\gamma >0$) 
$\alpha,\gamma $-heavy-pseudoparticle gap is 
$2\gamma |\mu|$ and $2(1+\gamma)\mu_0|H|+
\epsilon^0_{s,\gamma }(0)$ for $c$ and $s$, respectively. 

The gap expression $(123)-(124)$ involves the chemical
potential $\mu $, magnetic field $H$, and the zero-momentum
$\alpha,\gamma$ (with $\gamma >0$) band values. For constant 
values of $U$ the above quantities change as follows:
For fixed values of $S^s_z$, $|\mu|$ changes between
$|\mu |={\Delta_{MH}\over 2}$ for ${S^c_z\over N_a}\rightarrow 0$ 
(and $n\rightarrow 1$) and a maximum value,
$|\mu |={\Delta_{MH}\over 2}+4t$, for $S^c_z\rightarrow 
\pm{N_a\over 2}$ (and both $n\rightarrow 0$ and
$n\rightarrow 2$), where $\Delta_{MH}$ is the half-filled
Mott-Hubbard gap \cite{Lieb,Carmelo91}. The gap $\Delta_{MH}$
is both an increasing function of $U$ and of $|S^s_z|$
\cite{Carmelo91}.
On the other hand, for fixed values of $S^c_z$, $|H|$ 
changes between $|H|=0$ for ${S^s_z\over 
N_a}\rightarrow 0$ (and $m\rightarrow 
0$) and a maximum value, $|H|=H_c$, for $S^s_z\rightarrow 
\pm{N\over 2}$ (and $m\rightarrow\pm n$ and
$n\rightarrow 2$), where $H_c$ is the critical field
for onset of fully polarized ferromagnetism defined in Refs.
\cite{Carmelo91,Carmelo92c}. [Its expression presented
in these papers refers to the parameter-space sectors 
where the electronic density is such that $0<n<1$. The 
generalization of that expression to the sectors where 
the density is such that $1<n<2$ is obtained by 
replacing in it $n$ by $2-n$.] 
Finally, for fixed values of $S^c_z$ the band $\epsilon^0_{s,\gamma }(0)$
changes between $\epsilon^0_{s,\gamma }(0)=0$
for ${S^s_z\over N_a}\rightarrow 0$ and a minimum negative value. 
A more detailed study of the $U$, density, and magnetization 
dependence of the gap expression $(123)-(124)$ will be presented 
elsewhere.

Given an initial GS and a final sub-canonical ensemble we can write 

\begin{equation}
:{\hat{H}}: = {\hat{H}}_{0} + {\hat{H}}_{Landau} \, , 
\end{equation}
where ${\hat{H}}_{0}$ has eigenvalue $\omega_0$ 
and corresponds to the GS - GGS energy (a) and 
${\hat{H}}_{Landau}$ is normal-ordered relative
to the initial GS (with energy shifted by $\omega_0$) and is of the 
form

\begin{equation}
{\hat{H}}_{Landau} = \hat{H}^{(1)}_L + \hat{H}^{(2)}_L \, , 
\end{equation}
where

\begin{equation}
\hat{H}^{(1)}_L=\sum_{q,\alpha,\gamma}
\epsilon_{\alpha ,\gamma}(q):\hat{N}_{\alpha ,\gamma}(q): \, ,
\end{equation}

\begin{equation}
\epsilon_{\alpha ,0}(q) = \epsilon^{(0)}_{\alpha ,0}(q) 
- \epsilon^{(0)}_{\alpha ,0}(q_{F\alpha,0}) 
\, , \hspace{1cm}
\epsilon_{\alpha ,\gamma}(q) = \epsilon^{(0)}_{\alpha ,\gamma}(q) 
- \delta_{\alpha ,s}\epsilon^{(0)}_{\alpha ,\gamma}(0) \, ,
\end{equation}
and $\hat{H}^{(2)}_L$ is given in Eq. $(106)$. The Hamiltonian $(127)$
describes both the GS - GGS transition and Landau-liquid 
pseudoparticle - pseudohole excitations relative to the GGS. 
(The $c,0$ and $s,0$ pseudoparticle bands are shown 
in Figs. 7 and 8, respectively, of Ref \cite{Carmelo91}.)
Here $(127)$ are the Landau-liquid Hamiltonian terms 
which are {\it relevant} at low energy $(\omega-\omega_0)$. 
Therefore, the corresponding second-order Hamiltonian is 
suitable to study the physics at low positive energy above 
the gap, {\it i.e.}, small $(\omega -\omega_0)$ 
\cite{Carmelo96b,Carmelo96d}. 

In general, the different final GGS's of the Hilbert subspace 
where a given initial GS is transformed upon excitations
involving a small density of pseudoparticles have different 
energies $(123)-(124)$. This implies that the study of the 
quantum-liquid physics at energy-scale $\omega_0$ involves, in 
general, transitions to one sub-cannonical ensemble only. 
When two or several possible final GGS's have the same energy gap 
($\omega_0$) the physics involves the Hilbert subspace spanned by all
Hamiltonian eigenstates associated with the corresponding 
different final sub-canonical ensembles. 

For sub-canonical ensembles such that $S^{\alpha}_z\neq 0$ and 
characterized by small values of $N^h_{\alpha ,{-l_{\alpha}\over 
2}}/N^h_{\alpha }$ [where $l_{\alpha }$ is given in Eq. $(7)$], 
the gap expression $(123)$ leads directly to 

\begin{equation}
{\hat{H}}_{0} = 2|\mu|{\hat{N}}^h_{c,-{l_c\over 2}}+ 
2\mu_0|H|[{\hat{N}}^h_{s,-{l_s\over 2}}
+\sum_{\gamma =1}^{\infty}{\hat{N}}_{s,\gamma}] 
+ \sum_{\gamma =1}\epsilon^{(0)}_{s ,\gamma }(0)
{\hat{N}}_{\alpha ,\gamma} \, .
\end{equation}

There is a remarkable similarity between the general Hamiltonian 
$(126)$ and its version in the Hilbert subspace spanned by the states I 
\cite{Carmelo94,Carmelo96,Carmelo94b}. Moreover, in the case that 
the GGS is the GS it self or a GS differing from it 
by one or a few electrons the gap $(123)-(124)$ 
vanishes and the Hamiltonian 
$(126)$ reduces to the above low-energy Hamiltonian. In this case it 
refers to gapless GS - GS transitions which change the electron
and $\alpha ,0$ pseudoparticle numbers and to
Landau-liquid $\alpha ,0$ pseudoparticle-pseudohole 
excitations around the final GS.

The two-pseudoparticle forward-scattering phase shifts 
$\Phi_{\alpha,\gamma;\alpha',\gamma'}(q,q')$ defined by 
Eqs. $(99)-(103)$ and (B30)-(B40) and the  
velocities $(115)$ play an 
important role for the physics at energy scales around
$\omega_0$ \cite{Carmelo96b,Carmelo96d}.
[In the case of $\gamma=0$, the velocities $(115)$ are plotted 
in Fig. 9 of Ref. \cite{Carmelo91}.] 

In this section we have presented the expression of the
normal-ordered Hamiltonian and rapidity operators in the 
pseudoparticle basis. This confirms 
the consistency of the one-dimensional Landau-liquid 
theory which was shown to refer to this basis. The advantage 
of using the pseudoparticle basis is that (i) the problem 
becomes perturbative, {\it i.e.} from the point of view of the 
pseudoparticle interactions it is possible to classify 
which scatterings are {\it relevant} and (ii) we can describe
the relevant high-energy GS transitions because these
refer to a small density of excited pseudoparticles and
are, therefore, within the range of the perturbation theory.

For energies $\omega $ just above the set of energy values $\omega_0$ 
of the form given in Eqs. $(123)-(124)$ only the 
two-pseudoparticle interactions are relevant, as can be confirmed 
by finite-size studies at the corresponding critical point
\cite{Carmelo96d}.
Therefore, and for simplicity, we have presented in this paper only 
the two first Hamiltonian terms of expression $(104)$. However, Eqs. 
$(66)-(68)$ and $(83)$ contain {\it full} information about
all the remaining terms of higher-scattering order. 
All these many-pseudoparticle terms only describe
zero-momentum forward-scattering interactions.

\section{CONCLUDING REMARKS}

In this paper we have introduced a pseudohole and 
heavy-pseudoparticle representation for all the Hamiltonian
eigenstates of the Hubbard chain in a magnetic field
and chemical potential which is valid for all sectors of 
Hamiltonian symmetry. In this picture {\it all}
Hamiltonian eigenstates can be generated from a single 
reference vacuum, the half-filling and zero-magnetic-field
GS.

Our algebraic approach permits an operator analysis of the BA
solution and goes beyond that solution by generating
the non-LWS's and non-HWS's of the $\eta$-spin and
spin algebras. We find that all GS excitations 
can be divided into (a) a GS - GGS topological transition
which changes the $\alpha,\beta$ pseudohole and
$\alpha,\gamma$ heavy-pseudoparticle numbers and
(b) one pseudoparticle-pseudohole excitation around
the final GGS. It will be shown elsewhere 
\cite{Carmelo96b,Carmelo96d} that the quantum-liquid
physically important transitions are the GS - GGS
transitions which occur at energies $\omega_0$ of form 
$(124)$. Further, for low $(\omega -\omega_0)$ energies our
algebraic approach leads to the
construction of a GS normal-ordered Hamiltonian which has in the 
pseudoparticle basis a universal form involving 
only $k = 0$, forward-scattering pseudoparticle interaction terms.

A central result of our paper is that the perturbative character 
of the pseudoparticle basis also refers to high-energy states
{\it provided} that the density of excited $\alpha,\gamma$ 
pseudoparticles is small. This perturbative character
implies that the two-pseudoparticle Landau
$f$ functions and forward-scattering amplitudes are
finite, in contrast to the non-perturbative 
electronic representation, in which the two-electron 
forward-scattering amplitudes and vertices diverge.

The pseudoparticle algebra allows us to see immediately
the origin of the ``universal'' character
of the one-dimensional integrable quantum liquids
at all energy scales, which can be understood as a straightforward
generalization to pseudoparticles of Wilson's low-energy
renormalization group arguments: the pseudo-Fermi points 
of the pseudoparticle GS's and GGS's replace the particle 
Fermi points, and close to the pseudo-Fermi points only few types of
two-pseudoparticle scattering processes are relevant
for the low $(\omega -\omega_0)$ energy physics.

Concerning applications of our ideas to other
theoretical models and to problems in real materials,
we note that it is of considerable interest to determine the 
extent to which the algebraic structure and pseudoparticle perturbation
theory remain valid in systems which are {\it not} integrable
but which behave as electronic Luttinger liquids; an example
of such a system is the one-dimensional extended Hubbard model.
In terms of real materials, it is known that the present 
one-dimensional quantum systems provide useful (albeit idealized) 
descriptions of the physics of quasi-one-dimensional solids.

Finally, we turn to directions for further research. In
addition to some specific calculations mentioned above, our
immediate aims are to apply the concepts and techniques
developed here to study the finite-frequency 
spectral and transport properties of the Hubbard chain.
The occurrence of only zero-momentum pseudoparticle forward 
scattering combined with the infinite pseudohole- and 
pseudoparticle-number conservations laws can be used to 
extract important information on the finite-frequency 
behavior of correlation functions \cite{Carmelo96d}. 
We will show elsewhere \cite{Carmelo96b} that 
the $\alpha ,\gamma$ pseudoparticles are the transport carriers 
at all energy scales and couple to external potentials and 
that the topological GS - GGS transitions introduced in this 
paper give rise to the finite-frequency-conductivity absorption
edges. We emphasize that our pseudoparticle operator basis provides
a more suitable starting point for future 
investigations on that open problem. This is a subject of considerable
interest for the understanding of the mechanisms behind
the unusual spectral and transport properties of the
novel low-dimensional materials. 

\nonum
\section{ACKNOWLEDGMENTS}

We thank D. K. Campbell and J. M. B. Lopes dos Santos for 
stimulating discussions, C. S. I. C. (Spain) and J. N. I. C. T. 
(Portugal) for finantial support, and the hospitality of 
Instituto de Ciencia de Materiales (Madrid), where part of this
work was performed. N. M. R. P. thanks the Gulbenkian and
Luso-American Foundations for finantial support and the
University of Illinois for its hospitality.

\vfill
\eject
\appendix{RELATION TO THE USUAL BA NOTATION}

Our pseudoparticle operator representation required a suitable
choice of notation which is directly related to the BA notation
used by Takahashi \cite{Takahashi}. In this Appendix we relate 
the pseudoparticle notation to his notation.
In our $c,\gamma$ and $s,\gamma$ parameters below
we consider that $\gamma >0$ and $\gamma =0,1,2...$, 
respectively. 

The $N_{\alpha,\gamma}$ pseudoparticle numbers are related to the 
numbers $N$, $M'$, ${M'}_{n}$, and $M_{n}$ of Ref. \cite{Takahashi} 
as 

\begin{equation}
N_{c,0}\equiv N-2M' \, , \hspace{1cm} N_{c,\gamma}\equiv {M'}_{\gamma} 
\, , \hspace{1cm} N_{s,\gamma}\equiv M_{1+\gamma} \, . 
\end{equation}
Moreover, our onsite parameter $u={U\over 4t}$ equals the interaction $U$ of 
that reference. Considering
the discrete pseudomomentum $q_j$ of Eq. $(28)$, our rapidity
eigenvalues are related to the rapidity-real parts
$k_j$, ${\Lambda'}^{n}_{\alpha }$, and ${\Lambda}^{n}_{\alpha }$
of Ref. \cite{Takahashi} as follows

\begin{equation}
K(q_j)\equiv k_j \, , \hspace{1cm} R_{c,\gamma}(q_j)\equiv 
{{\Lambda'}^{\gamma}_j\over u} 
\, , \hspace{1cm} R_{s,\gamma}(q_j)\equiv {\Lambda^{1+\gamma}_j\over u}  
\, . 
\end{equation}
Finally, the relation of the pseudoparticle quantum
numbers of Eq. $(28)$ to the Takahashi quantum numbers $I_j$, 
${J'}^n_{\alpha}$, and $J^n_{\alpha}$ is

\begin{equation}
I_j^{c,0}\equiv I_j \, , \hspace{1cm} I_j^{c,\gamma}\equiv 
{J'}^{\gamma}_j \, , \hspace{1cm} I_j^{s,\gamma}\equiv J^{1+\gamma}_j  
\, . 
\end{equation}

\vfill
\eject
\appendix{NORMAL-ORDERED RAPIDITY OPERATOR EXPRESSIONS}

Following the discussion of Sec. V, the perturbative character of 
the system implies the equivalence between expanding in the 
pseudoparticle scattering order and/or in the pseudomomentum 
deviations $(95)$. In this Appendix we give a short 
description of the calculation of the normal-ordered operator 
expansion for the pseudoparticle rapidities $(97)$. We focus our 
study on the evaluation of the functions $X_{\alpha,\gamma}^{(1)}(q)$ 
and $X_{\alpha,\gamma}^{(2)}(q)$ and associate functions
$Q_{\alpha,\gamma}^{(1)}(q)$ and $Q_{\alpha,\gamma}^{(2)}(q)$ 
(eigenvalues of the operators $\hat{X}_{\alpha}^{(1)}(q)$ and 
$\hat{X}_{\alpha}^{(2)}(q)$ and $\hat{Q}_{\alpha}^{(1)}(q)$ and 
$\hat{Q}_{\alpha}^{(2)}(q)$, respectively).

We evaluate here the first-order and second-order
terms of the eigenvalues of the operator $(97)$. Equation $(86)$ then 
allows the straightforward calculation of the corresponding 
operator expressions.

In the thermodynamic limit, Eqs. $(66)-(68)$ lead to
the following equations 
 
\begin{eqnarray}
K(q) & = & q - {1\over \pi}\sum_{\gamma =0}
\int_{q^{(-)}_{s,\gamma}}^{q^{(+)}_{s,\gamma}}dq'
N_{s,\gamma}(q')\tan^{-1}\Bigl({R_{c,0}(q)-R_{s,\gamma}(q')
\over 1+\gamma}\Bigr)
\nonumber \\
& - & {1\over \pi}\sum_{\gamma =1}
\int_{q^{(-)}_{c,\gamma}}^{q^{(+)}_{c,\gamma}}dq'
N_{c,\gamma}(q')
\tan^{-1}\Bigl({R_{c,0}(q)-R_{c,\gamma}(q')
\over \gamma}\Bigr) \, ,
\end{eqnarray}

\begin{eqnarray}
2 Re \sin^{-1}\Bigl(u(R_{c,\gamma}(q)-i\gamma)\Bigr) 
& = & q + {1\over\pi}\int_{q^{(-)}_{c,0}}^{q^{(+)}_{c,0}}dq'
N_{c,0}(q')\tan^{-1}\Bigl({R_{c,\gamma}(q)-R_{c,0}(q')\over 
\gamma}\Bigr)
\nonumber \\
& + & {1\over 2\pi}\sum_{\gamma' =1}
\int_{q^{(-)}_{c,\gamma'}}^{q^{(+)}_{c,\gamma'}}dq'
N_{c,\gamma'}(q')
\Theta_{\gamma,\gamma'}\Bigl(R_{c,\gamma}(q)-R_{c,\gamma'}(q')\Bigr) \, ,
\end{eqnarray}
and 

\begin{eqnarray}
0 & = & q - {1\over\pi}
\int_{q^{(-)}_{c,0}}^{q^{(+)}_{c,0}}dq'
N_{c,0}(q')\tan^{-1}\Bigl({R_{s,\gamma}(q)-R_{c,0}(q')
\over 1+\gamma}\Bigr)\nonumber \\
& + & {1\over 2\pi}\sum_{\gamma' =0}
\int_{q^{(-)}_{s,\gamma'}}^{q^{(+)}_{s,\gamma'}}dq'
N_{s,\gamma'}(q')\Theta_{1+\gamma, 1+\gamma'}
\Bigl(R_{s,\gamma}(q)-R_{s,\gamma'}(q')\Bigr) \, .
\end{eqnarray}

We start by considering the GGS eigenstate rapidities of 
Eqs. $(72)$ and $(73)$. If we insert in Eqs. $(66)-(68)$ the 
GGS distribution $(75)$ after some algebra we find that the 
functions of the rhs of Eqs. $(77)-(78)$ defining the inverse
of these GGS rapidities are solutions of the following integral 
equations

\begin{equation}
2\pi\rho_{c,0}(k) = 1
+ {\cos k\over u}
\int_{r^{(-)}_{s,0}}^{r^{(+)}_{s,0}}dr 
{2\pi\rho_{s,0}(r)\over \pi 
\left[1+(r - {\sin k\over u})^2\right]} \, ,
\end{equation}

\begin{equation}
2\pi\rho_{c,\gamma}(r) = 2 Re \Bigl({u\over
\sqrt{1 - u^2[r - i\gamma]^2}}\Bigl)
- \int_{Q^{(-)}}^{Q^{(+)}}dk {2\pi\rho_{c,0}(k)\over
\pi\gamma\left[1+({{\sin k\over u} -r\over\gamma})^2\right]} \, ,
\end{equation}
and

\begin{eqnarray}
2\pi\rho_{s,\gamma}(r) & = & 
\int_{Q^{(-)}}^{Q^{(+)}}dk {2\pi\rho_{c,0}(k)\over
\pi(1+\gamma )\left[1+({{\sin k\over u} -r\over 1+\gamma })^2\right]} 
\nonumber \\
& - & \int_{r^{(-)}_{s,0}}^{r^{(+)}_{s,0}}dr'{\Theta^{[1]}_{1,1+\gamma }
\Bigl(r -r'\Bigl)\over 2\pi}2\pi\rho_{s,0}(r') \, ,
\end{eqnarray}
where

\begin{eqnarray}
\Theta^{[1]}_{\gamma,\gamma'}\Bigl(x\Bigl) & = &
\Theta^{[1]}_{\gamma',\gamma}\Bigl(x\Bigl) =
{d\Theta_{\gamma,\gamma'}\Bigl(x\Bigl)\over dx} \nonumber \\
& = & \delta_{\gamma ,\gamma'}\{{1\over \gamma[1+({x\over 
2\gamma})^2]}+ \sum_{l=1}^{\gamma -1}{2\over l[1+({x\over 2l})^2]}\}
\nonumber \\ 
& + & (1-\delta_{\gamma ,\gamma'})\{
{2\over |\gamma-\gamma'|[1+({x\over |\gamma-\gamma'|})^2]}
+ {2\over (\gamma+\gamma')[1+({x\over \gamma+\gamma'})^2]}\nonumber \\
& + & \sum_{l=1}^{{\gamma+\gamma'-|\gamma-\gamma'|\over 2}
-1}{4\over (|\gamma-\gamma'|+2l)[1+({x\over |\gamma-\gamma'|+2l})^2]}\} \, ,
\end{eqnarray}
is the derivative of the function $(69)$ and the parameters 
$Q^{(\pm)}$ and $r^{(\pm)}_{\alpha,\gamma }$ are defined 
combining Eqs. $(77)-(78)$ and $(79)$. We have omitted in the
rhs of Eqs. (B4)-(B6) terms associated with contributions of order 
$j>2$ in the density of excited pseudoparticles. These terms
do not contribute to the quantities to be evaluated in this
paper, as we discuss below.

Let us now consider small deviations from a GGS or GS.
The eigenvalue form of Eq. $(96)$ is  

\begin{equation}
\delta K(q) = K^{(0)}(q + 
\delta X_{c,0}(q)) - K^{(0)}(q) \, ,
\end{equation}
for $\alpha =c$ and $\gamma =0$ and

\begin{equation}
\delta R_{\alpha,\gamma}(q) = R_{\alpha,\gamma}^0(q + 
\delta X_{\alpha,\gamma}(q)) - R_{\alpha,\gamma}^0(q) \, ,
\end{equation}
for all remaining values of the quantum numbers $\alpha $
and $\gamma $. Here $\delta K(q)$, $\delta R_{\alpha,\gamma}(q)$, and 
$\delta X_{\alpha,\gamma}(q)$ are the eigenvalues of the operators 
$:\hat{K}(q):$, $:\hat{R}_{\alpha,\gamma}(q):$, and 
$:\hat{X}_{\alpha,\gamma}(q):$, respectively. From Eq. $(97)$
$\delta X_{\alpha,\gamma}(q)$ can be written as

\begin{equation}
\delta X_{\alpha,\gamma}(q) = X_{\alpha,\gamma}^{(1)}(q) + 
X_{\alpha,\gamma}^{(2)}(q)
+ ... \, , 
\end{equation}
where $X_{\alpha,\gamma}^{(i)}(q)$ is the eigenvalue of the operator 
$\hat{X}_{\alpha,\gamma}^{(i)}(q)$. Expanding the 
$\delta R_{\alpha,\gamma}(q)$ expressions (B8) and (B9) we find

\begin{equation}
K(q) = \sum_{i=0}^{\infty} K^{(i)}(q) \, ,
\end{equation}
and

\begin{equation}
R_{\alpha,\gamma}(q) = \sum_{i=0}^{\infty} R_{\alpha,\gamma}^{(i)}(q) \, ,
\end{equation}
respectively, [and $\delta K(q) = \sum_{i=1}^{\infty} K^{(i)}(q)$ and
$\delta R_{\alpha,\gamma}(q)=\sum_{i=1}^{\infty}
R_{\alpha,\gamma}^{(i)}(q)$] where the zero-order GGS
functions $K^{(0)}(q)$ and $R_{\alpha,\gamma}^{(0)}(q)$ are 
defined by Eqs. $(77)-(78)$. From the resulting equations we can
obtain all derivatives of the GGS functions $K^{(0)}(q)$ and
$R_{\alpha,\gamma}^{(0)}(q)$ with respect to $q$. The terms of the 
rhs of Eqs. (B11) and (B12) involve these derivatives. For instance,
the first-order and second-order terms read

\begin{equation}
K^{(1)}(q) =  {d K^{(0)}(q)\over {dq}} 
X_{c,0}^{(1)}(q) \, ,
\end{equation}

\begin{equation}
R_{\alpha,\gamma}^{(1)}(q) =  {d R_{\alpha,\gamma}^{(0)}(q)\over {dq}} 
X_{\alpha,\gamma}^{(1)}(q) \, ,
\end{equation}
and

\begin{equation}
K^{(2)}(q) =  {d K^{(0)}(q)\over {dq}} 
X_{c,0}^{(2)}(q) + 
{1\over 2}{d^2 K^{(0)}(q)\over 
{dq^2}}[X_{c,0}^{(1)}(q)]^2 \, ,
\end{equation}

\begin{equation}
R_{\alpha,\gamma}^{(2)}(q) =  {d R_{\alpha,\gamma}^{(0)}(q)\over {dq}} 
X_{\alpha,\gamma}^{(2)}(q) + 
{1\over 2}{d^2 R_{\alpha,\gamma}^{(0)}(q)\over 
{dq^2}}[X_{\alpha,\gamma}^{(1)}(q)]^2 \, ,
\end{equation}
respectively, and involve the first and second derivatives.

From Eqs. (B1)-(B3) [with $N_{\alpha,\gamma}(q')$ given
by the GGS distribution $(75)$] we find that
the first derivatives $d K_{\alpha}^{(0)}(q)\over 
{dq}$ and $d R_{\alpha,\gamma}^{(0)}(q)\over 
{dq}$ can be expressed in terms of the functions 
(B4)-(B6) as follows 

\begin{equation}
{d K^{(0)}(q)\over {dq}} ={1\over 2\pi\rho_{c,0}(K^{(0)}(q))}
\, ,
\end{equation}
and

\begin{equation}
{d R^{(0)}_{\alpha,\gamma}(q)\over {dq}} =
{1\over 2\pi\rho_{c,\gamma}(R^{(0)}_{\alpha,\gamma}(q))}
\, .
\end{equation}
The second derivatives $d^2 K^{(0)}(q)\over {dq^2}$ and
$d^2 R_{\alpha,\gamma}^{(0)}(q)\over {dq^2}$ then read

\begin{equation}
{d^2 K^{(0)}(q)\over {dq^2}} =
- {1\over [2\pi\rho_{c,0}(K^{(0)}(q))]^3}
2\pi{d \rho_{c,0}(k) \over {dk}}|_{k=K^{(0)}(q)} \, ,
\end{equation}
and

\begin{equation}
{d^2 R_{\alpha,\gamma}^{(0)}(q)\over {dq^2}} = 
-{1\over [2\pi\rho_{c,\gamma}(R^{(0)}_{\alpha,\gamma}(q))]^3}
2\pi{d \rho_{\alpha,\gamma}(r) \over 
{dr}}|_{r=R_{\alpha,\gamma}^{(0)}(q)} \, .
\end{equation}

By introducing both the distributions

\begin{equation}
N_{\alpha,\gamma}(q) = N_{\alpha,\gamma}^{(0)}(q) + 
\delta N_{\alpha,\gamma}(q) \, ,
\end{equation} 
and the first-order and second-order functions (B13)-(B16) 
into Eqs. (B1)-(B3), we find after expanding to second order that for
$j=1$ and $j=2$ the functions $X^{(j)}_{\alpha,\gamma}(q)$ can 
be written as 

\begin{equation}
\hat{X}_{\alpha,\gamma}^{(j)}(q) = \hat{Q}_{\alpha,\gamma}^{(j)}(q) + 
\hat{Y}_{\alpha,\gamma}^{(j)}(q) \, ,
\end{equation}
where $\hat{Y}_{\alpha,\gamma}^{(j)}(q)$ is even in $q$

\begin{equation}
\hat{Y}_{\alpha,\gamma}^{(j)}(q) = \hat{Y}_{\alpha,\gamma}^{(j)}(-q) \, ,
\end{equation}
and does not contribute to the physical quantities to second
scattering order (and to second order in the density of excited
pseudoparticles). Up to this order only 
$\hat{Q}_{\alpha,\gamma}^{(j)}(q)$ contributes. 

To derive this result we have expanded the expression
for the even functions $Y^{(j)}_{\alpha,\gamma}(q)$
for $j=1$ and $j=2$ to second order in the density of excited 
pseudoparticles. Following the perturbative character of the quantum
liquid in the pseudoparticle basis, the obtained expression 
is exact up to $j=2$ pseudoparticle scattering order.
Moreover, we find that the even function $Y^{(1)}_{\alpha,\gamma}(q)$ 
does not contribute to the physical quantities up
to that order and can, therefore, be omitted.
 
Let us introduce the functions $\bar{Q}_{\alpha,\gamma}^{(1)}(r)$
such that

\begin{equation}
Q_{\alpha,\gamma}^{(1)}(q) = 
\bar{Q}_{\alpha,\gamma}^{(1)}(R^{(0)}_{\alpha,\gamma}(q)) \, ,
\end{equation}
for all values of $\alpha$ and $\gamma$. Note that following
Eq. $(64)$

\begin{equation}
R^{(0)}_{c,0}(q) = {\sin K^{(0)}(q)\over u} \, .
\end{equation}
It is also useful to define the function
$\widetilde{Q}^{(1)}(k)$ such that

\begin{equation}
\widetilde{Q}^{(1)}(k) =
{\bar{Q}}_{c,0}^{(1)}({\sin k\over u}) \, .
\end{equation}

By introducing both the distributions (B21) and the 
first-order functions defined in Eq. (B24) into 
Eqs. (B1)-(B3), we find after expanding to first order that
the functions $\bar{Q}_{\alpha,\gamma}^{(1)}(r)$ are 
defined by the following system of coupled integral
equations

\begin{eqnarray}
{\bar{Q}}_{c,0}^{(1)}(r) & = &
- \sum_{\gamma =0}\int_{q_{s,\gamma}^{(-)}}^{q_{s,\gamma}^{(+)}}dq 
\delta N_{s,\gamma}(q){1\over \pi}
\tan^{-1}\Bigl({r - R_{s,\gamma}^{(0)}(q)\over 1+\gamma}\Bigl)
\nonumber \\
& - & \sum_{\gamma =1}\int_{q_{c,\gamma}^{(-)}}^{q_{c,\gamma}^{(+)}}dq 
\delta N_{c,\gamma}(q){1\over \pi}
\tan^{-1}\Bigl({r - R_{c,\gamma}^{(0)}(q)\over \gamma}\Bigl)
+ \int_{r^{(-)}_{s,0}}^{r^{(+)}_{s,0}}
dr'{\bar{Q}_{s,0}^{(1)}(r')\over 
\pi\left[1 + (r -r')^2\right]} \, ,
\end{eqnarray}

\begin{eqnarray}
{\bar{Q}}_{c,\gamma}^{(1)}(r) & = &
\int_{q_{c,0}^{(-)}}^{q_{c,0}^{(+)}}dq
\delta N_{c,0}(q){1\over \pi}
\tan^{-1}\Bigl({r - R_{c,0}^{(0)}(q)\over \gamma}\Bigl)
\nonumber \\
& + & \sum_{\gamma' =1}\int_{q_{c,\gamma'}^{(-)}}^{q_{c,\gamma'}^{(+)}}dq 
\delta N_{c,\gamma'}(q){1\over 2\pi}\Theta_{\gamma,\gamma'}
\Bigl(r - R_{c,\gamma'}^{(0)}(q)\Bigl)
- \int_{r^{(-)}_{c,0}}^{r^{(+)}_{c,0}}
dr'{\bar{Q}_{c,0}^{(1)}(r')\over 
\pi\gamma\left[1 + ({r -r'\over\gamma })^2\right]} \, ,
\end{eqnarray}
and

\begin{eqnarray}
{\bar{Q}}_{s,\gamma}^{(1)}(r) & = &
- \int_{q_{c,0}^{(-)}}^{q_{c,0}^{(+)}}dq
\delta N_{c,0}(q){1\over \pi}
\tan^{-1}\Bigl({r - R_{c,0}^{(0)}(q)\over 1+\gamma}\Bigl)
\nonumber \\
& + & \sum_{\gamma' =0}
\int_{q_{s,\gamma'}^{(-)}}^{q_{s,\gamma'}^{(+)}}dq 
\delta N_{s,\gamma'}(q){1\over 2\pi}\Theta_{1+\gamma,1+\gamma'}
\Bigl(r - R_{s,\gamma'}^{(0)}(q)\Bigl)
+ \int_{r^{(-)}_{c,0}}^{r^{(+)}_{c,0}}
dr'{\bar{Q}_{c,0}^{(1)}(r')\over 
\pi(1+\gamma)\left[1 + ({r -r'\over 1+\gamma })^2\right]} 
\nonumber \\
& - & \int_{r^{(-)}_{s,0}}^{r^{(+)}_{s,0}}
dr'\bar{Q}_{s,0}^{(1)}(r'){1\over 
2\pi}\Theta^{[1]}_{1+\gamma,1}
\Bigl(r - r'\Bigl) \, .
\end{eqnarray}

[The use of Eq. (B27) in Eqs. (B28) and (B29) allows the 
expression of both ${\bar{Q}}_{c,\gamma}^{(1)}(r)$
and ${\bar{Q}}_{s,\gamma}^{(1)}(r)$ in terms of
free terms and integrals involving $\bar{Q}_{s,0}^{(1)}(r)$.]
Combining Eqs. (B27)-(B29) with Eqs. (B24) and (B26) leads to 
Eq. $(98)$ with the phase shifts defined below. Note that at 
first order we can either consider the function 
$\hat{X}_{\alpha,\gamma}^{(1)}(q)$ or the associate function 
$\hat{Q}_{\alpha,\gamma}^{(1)}(q)$ of Eq. (B22). 
Both functions are of the form $(98)$. However,
the general expressions for the phase shifts associated
with the function $\hat{X}_{\alpha,\gamma}^{(1)}(q)$
of Eq. $(98)$ have extra terms. These arise from the
function $\hat{Y}_{\alpha,\gamma}^{(1)}(q)$ of the rhs
of Eq. (B22). If we expand the physical quantities involving 
the phase shifts to first (and second) order in the density 
of excited pseudoparticles these extra terms lead to vanishing 
contributions. For instance, expression $(98)$ is identical if 
we use in it either choice for the phase shift expressions.
For simplicity, we omit here the phase-shift extra terms 
associated with the function $\hat{Y}_{\alpha,\gamma}^{(1)}(q)$.
We find that the phase shifts associated with the function
$\hat{Q}_{\alpha,\gamma}^{(1)}(q)$ are defined by the following 
coupled integral equations

\begin{equation}
\bar{\Phi }_{c,0;c,0}\left(r,r'\right) = 
{1\over{\pi}}\int_{r_{s,0}^{(-)}}^{r_{s,0}^{(+)}}
dr''{\bar{\Phi }_{s,0;c,0}\left(r'',r'\right) 
\over {1+(r-r'')^2}} \, ,
\end{equation}

\begin{equation}
\bar{\Phi }_{c,0;c,\gamma}\left(r,r'\right) = 
-{1\over{\pi}}\tan^{-1}({r-r'\over \gamma}) +
{1\over{\pi}}\int_{r_{s,0}^{(-)}}^{r_{s,0}^{(+)}}
dr''{\bar{\Phi }_{s,0;c,\gamma}\left(r'',r'\right) 
\over {1+(r-r'')^2}} \, ,
\end{equation}

\begin{equation}
\bar{\Phi }_{c,0;s,\gamma}\left(r,r'\right) = 
-{1\over{\pi}}\tan^{-1}({r-r'\over 1+\gamma}) + {1\over{\pi}}
\int_{r_{s,0}^{(-)}}^{r_{s,0}^{(+)}}
dr''{\bar{\Phi }_{s,0;s,\gamma}\left(r'',r'\right) 
\over {1+(r-r'')^2}} \, ,
\end{equation}

\begin{equation}
{\bar{\Phi }}_{c,\gamma;c,0}\left(r,r'\right) = 
{1\over{\pi}}\tan^{-1}({r-r'\over {\gamma}}) -
{1\over{\pi}}\int_{r_{c,0}^{(-)}}^{r_{c,0}^{(+)}}
dr''{{\bar{\Phi }}_{c,0;c,0}\left(r'',r'\right) 
\over {\gamma[1+({r-r''\over {\gamma}})^2]}} \, ,
\end{equation}

\begin{equation}
\bar{\Phi }_{c,\gamma;c,\gamma'}\left(r,r'\right) = 
{1\over{2\pi}}\Theta_{\gamma,\gamma'}\Bigl(r-r'\Bigr)
- {1\over{\pi}}\int_{r_{c,0}^{(-)}}^{r_{c,0}^{(+)}}
dr''{\bar{\Phi }_{c,0;c,\gamma'}\left(r'',r'\right) 
\over {\gamma[1+({r-r''\over\gamma})^2]}} \, ,
\end{equation}

\begin{equation}
\bar{\Phi }_{c,\gamma;s,\gamma'}\left(r,r'\right) = 
- {1\over{\pi}}\int_{r_{c,0}^{(-)}}^{r_{c,0}^{(+)}}
dr''{\bar{\Phi }_{c,0;s,\gamma'}\left(r'',r'\right) 
\over {\gamma[1+({r-r''\over\gamma})^2]}} \, ,
\end{equation}

\begin{eqnarray}
{\bar{\Phi }}_{s,\gamma;c,0}\left(r,r'\right) & = &
-{1\over{\pi}}\tan^{-1}({r-r'\over {1+\gamma}}) + 
{1\over{\pi}}\int_{r_{c,0}^{(-)}}^{r_{c,0}^{(+)}}
dr''{{\bar{\Phi }}_{c,0;c;0}\left(r'',r'\right) 
\over {(1+\gamma)[1+({r-r''\over {1+\gamma}})^2]}} 
\nonumber \\
& - & \int_{r_{s,0}^{(-)}}^{r_{s,0}^{(+)}}
dr''{\bar{\Phi }}_{s,0;c,0}\left(r'',r'\right)
{\Theta^{[1]}_{1+\gamma,1}\Bigl(r-r''\Bigr)\over{2\pi}} \, ,
\end{eqnarray}

\begin{eqnarray}
{\bar{\Phi }}_{s,\gamma;c,\gamma'}\left(r,r'\right) & = &
{1\over{\pi}}\int_{r_{c,0}^{(-)}}^{r_{c,0}^{(+)}}
dr''{{\bar{\Phi }}_{c,0;c;\gamma'}\left(r'',r'\right) 
\over {(1+\gamma)[1+({r-r''\over {1+\gamma}})^2]}} 
\nonumber \\
& - & \int_{r_{s,0}^{(-)}}^{r_{s,0}^{(+)}}
dr''{\bar{\Phi }}_{s,0;c,\gamma'}\left(r'',r'\right)
{\Theta^{[1]}_{1+\gamma,1}\Bigl(r-r''\Bigr)\over {2\pi}} \, ,
\end{eqnarray}

\begin{eqnarray}
{\bar{\Phi }}_{s,\gamma;s,\gamma'}\left(r,r'\right) & = &
{\Theta_{1+\gamma,1+\gamma'}\Bigl(r-r'\Bigr)\over{2\pi}} +
{1\over{\pi}}\int_{r_{c,0}^{(-)}}^{r_{c,0}^{(+)}}
dr''{{\bar{\Phi }}_{c,0;s;\gamma'}\left(r'',r'\right) 
\over {(1+\gamma)[1+({r-r''\over {1+\gamma}})^2]}} 
\nonumber \\
& - & \int_{r_{s,0}^{(-)}}^{r_{s,0}^{(+)}}
dr''{\bar{\Phi }}_{s,0;s,\gamma'}\left(r'',r'\right)
{\Theta^{[1]}_{1+\gamma,1}\Bigl(r-r''\Bigr)\over{2\pi}} \, .
\end{eqnarray}

For $\gamma =0$ Eqs. (B36) and (B38) can be rewritten as

\begin{equation}
\bar{\Phi }_{s,0;c,0}\left(r,r'\right) = 
-{1\over{\pi}}\tan^{-1}(r-r') + 
\int_{r_{s,0}^{(-)}}^{r_{s,0}^{(+)}}
dr''G(r,r''){\bar{\Phi }}_{s,0;c,0}\left(r'',r'\right) \, ,
\end{equation}
and

\begin{eqnarray}
\bar{\Phi }_{s,0;s,0}\left(r,r'\right) & =
& {1\over{\pi}}\tan^{-1}({r-r'\over{2}}) -
{1\over{\pi^2}}\int_{r_{c,0}^{(-)}}^{r_{c,0}^{(+)}} 
dr''{\tan^{-1} 
(r''-r')\over{1+(r-r'')^2}}
\nonumber \\
& + & \int_{r_{s,0}^{(-)}}^{r_{s,0}^{(+)}}
dr''G(r,r''){\bar{\Phi }}_{s,0;s,0}\left(r'',r'\right) \, ,
\end{eqnarray}
and the kernel $G(r,r')$ reads \cite{Carmelo92b}

\begin{equation}
G(r,r') = - {1\over{2\pi}}\left[{1\over{1+((r-r')/2)^2}}\right]
\left[1 - {1\over 2}
\left(t(r)+t(r')+{{l(r)-l(r')}\over{r-r'}}\right)\right] \, ,
\end{equation}
with

\begin{equation}
t(r) = {1\over{\pi}}\left[\tan^{-1}(r + r_{c,0}^{(+)}) 
- \tan^{-1}(r + r_{c,0}^{(-)})\right]\, ,
\end{equation}
and

\begin{equation}
l(r) = {1\over{\pi}}\left[
\ln (1+(r + r_{c,0}^{(+)})^2) -  
\ln (1+(r + r_{c,0}^{(-)})^2)\right] \, .
\end{equation}

In order to evaluate the second-order functions
$Q_{\alpha,\gamma}^{(2)}(q)$ of the rhs of Eq. (B22) for $j=2$ 
[see also Eq. (B10)] we introduce the functions (B13)-(B16)  
and distributions (B21) in Eqs. (B1)-(B3). 
Expanding to second order we find after some algebra
 
\begin{equation}
Q_{\alpha,\gamma}^{(2)}(q) = Q_{\alpha,\gamma}^{(2,*)}(q) + 
{1\over 2} {d\over {dq}}[[Q_{\alpha,\gamma}^{(1)}(q)]^2] \, , 
\end{equation}
where

\begin{equation}
Q_{\alpha,\gamma}^{(2,*)}(q) = 
\widetilde{Q}_{\alpha,\gamma}^{(2,*)}(R_{\alpha,\gamma}^{(0)}(q)) \, .
\end{equation}
It is also useful to define the function
$\widetilde{Q}^{(2,*)}(k)$ such that

\begin{equation}
\widetilde{Q}^{(2,*)}(k) =
{\bar{Q}}_{c,0}^{(2,*)}({\sin k\over u}) \, .
\end{equation}
The functions $\bar{Q}_{\alpha,\gamma}^{(2,*)}(r)$ are 
defined by the following system of coupled integral equations

\begin{eqnarray}
\bar{Q}_{c,0}^{(2,*)}(r) & = & \sum_{\gamma =0}
\int_{q_{s,\gamma}^{(-)}}^{q_{s,\gamma}^{(+)}}dq \delta N_{s,\gamma}(q)
{Q_{s,\gamma}^{(1)}(q)\over 
2\pi\rho_{s,\gamma}\Bigl(R^{(0)}_{s,\gamma}(q)\Bigl)}
{1\over \pi (1+\gamma)\left[1 + ({r - R^{(0)}_{s,\gamma}(q)
\over 1+\gamma})^2\right]}
\nonumber\\
& + & \sum_{\gamma =1}\int_{q_{c,\gamma}^{(-)}}^{q_{c,\gamma}^{(+)}}dq 
\delta N_{c,\gamma}(q){Q_{c,\gamma}^{(1)}(q)\over 
2\pi\rho_{c,\gamma}\Bigl(R^{(0)}_{c,\gamma}(q)\Bigl)}
{1\over \pi\gamma\left[1 + ({r - R^{(0)}_{c,\gamma}(q)
\over\gamma})^2\right]}
\nonumber\\
& + & \sum_{\gamma =0}{\theta (N_{s,\gamma})\over 2\pi\rho_{s,\gamma}
\Bigl(r_{s,\gamma}\Bigl)}\sum_{j=\pm 1}
{[Q_{s,\gamma}^{(1)}(jq_{Fs,\gamma})]^2
\over 2\pi (1+\gamma )\left[1 + ({r-jr_{s,\gamma}
\over 1+\gamma})^2\right]}
\nonumber\\
& - & \sum_{\gamma =1}{\theta (N_{c,\gamma})\over 2\pi\rho_{c,\gamma}
\Bigl(r_{c,\gamma}\Bigl)}\sum_{j=\pm 1}
{[Q_{c,\gamma}^{(1)}(jq_{Fc,\gamma})]^2
\over 2\pi\gamma\left[1 + ({r-jr_{c,\gamma}
\over\gamma})^2\right]} 
\nonumber\\
& + & \int_{r^{(-)}_{s,0}}^{r^{(+)}_{s,0}}
dr'{\bar{Q}_{s,0}^{(2,*)}(r')\over 
\pi\left[1 + (r -r')^2\right]} \, ,
\end{eqnarray}

\begin{eqnarray}
\bar{Q}_{c,\gamma}^{(2,*)}(r) & = & 
- \int_{q_{c,0}^{(-)}}^{q_{c,0}^{(+)}}dq 
\delta N_{c,0}(q){\cos K^{(0)}(q)\over u}
{Q_{c,0}^{(1)}(q)\over 
2\pi\rho_{c,0}\Bigl(k^{(0)}(q)\Bigl)}
{1\over \pi\gamma\left[1 + ({r - R^{(0)}_{c,0}(q)
\over\gamma})^2\right]}
\nonumber\\
& - & \sum_{\gamma' =1}\int_{q_{c,\gamma'}^{(-)}}^{q_{c,\gamma'}^{(+)}}dq 
\delta N_{c,\gamma'}(q){Q_{c,\gamma'}^{(1)}(q)\over 
2\pi\rho_{c,\gamma'}\Bigl(R^{(0)}_{c,\gamma'}(q)\Bigl)}
{1\over 2\pi}\Theta^{[1]}_{\gamma ,\gamma'}\Bigl(r - 
R^{(0)}_{c,\gamma'}(q)\Bigl)
\nonumber\\
& - & {\cos Q\over u}{1\over 2\pi\rho_{c,0}
\Bigl(Q\Bigl)}\sum_{j=\pm 1}
{[Q_{c,0}^{(1)}(jq_{Fc,0})]^2
\over 2\pi\gamma\left[1 + ({r-jr_{c,0}
\over\gamma})^2\right]} 
\nonumber\\
& + & \sum_{\gamma' =1}{\theta (N_{c,\gamma'})\over 2\pi\rho_{c,\gamma'}
\Bigl(r_{c,\gamma'}\Bigl)}{1\over 2}\sum_{j=\pm 1}
[Q_{c,\gamma'}^{(1)}(jq_{Fc,\gamma'})]^2
{1\over 2\pi}\Theta^{[1]}_{\gamma ,\gamma'}
\Bigl(r-jr_{c,\gamma'}\Bigl)
\nonumber\\
& - & \int_{r^{(-)}_{c,0}}^{r^{(+)}_{c,0}}
dr'{\bar{Q}_{c,0}^{(2,*)}(r')\over 
\pi\gamma\left[1 + ({r -r'\over\gamma})^2\right]} \, ,
\end{eqnarray}
and

\begin{eqnarray}
\bar{Q}_{s,\gamma}^{(2,*)}(r) & = & 
\int_{q_{c,0}^{(-)}}^{q_{c,0}^{(+)}}dq 
\delta N_{c,0}(q){\cos K^{(0)}(q)\over u}
{Q_{c,0}^{(1)}(q)\over 
2\pi\rho_{c,0}\Bigl(k^{(0)}(q)\Bigl)}
{1\over \pi (1+\gamma )\left[1 + ({r - R^{(0)}_{c,0}(q)
\over 1+\gamma})^2\right]}
\nonumber\\
& - & \sum_{\gamma' =0}\int_{q_{s,\gamma'}^{(-)}}^{q_{s,\gamma'}^{(+)}}dq 
\delta N_{s,\gamma'}(q){Q_{s,\gamma'}^{(1)}(q)\over 
2\pi\rho_{s,\gamma'}\Bigl(R^{(0)}_{s,\gamma'}(q)\Bigl)}
{1\over 2\pi}\Theta^{[1]}_{1+\gamma ,1+\gamma'}\Bigl(r - 
R^{(0)}_{s,\gamma'}(q)\Bigl)
\nonumber\\
& + & {\cos Q\over u}{1\over 2\pi\rho_{c,0}
\Bigl(Q\Bigl)}\sum_{j=\pm 1}
{[Q_{c,0}^{(1)}(jq_{Fc,0})]^2
\over 2\pi (1+\gamma)\left[1 + ({r-jr_{c,0}
\over 1+\gamma})^2\right]} 
\nonumber\\
& - & \sum_{\gamma' =0}{\theta (N_{s,\gamma'})\over 2\pi\rho_{s,\gamma'}
\Bigl(r_{s,\gamma'}\Bigl)}{1\over 2}\sum_{j=\pm 1}
[Q_{s,\gamma'}^{(1)}(jq_{Fs,\gamma'})]^2
{1\over 2\pi}\Theta^{[1]}_{1+\gamma ,1+\gamma'}
\Bigl(r-jr_{s,\gamma'}\Bigl)
\nonumber\\
& + & \int_{r^{(-)}_{c,0}}^{r^{(+)}_{c,0}}
dr'{\bar{Q}_{c,0}^{(2,*)}(r')\over 
\pi(1+\gamma)\left[1 + ({r -r'\over 1+\gamma })^2\right]} 
\nonumber \\
& - & \int_{r^{(-)}_{s,0}}^{r^{(+)}_{s,0}}
dr'\bar{Q}_{s,0}^{(2,*)}(r'){1\over 
2\pi}\Theta^{[1]}_{1+\gamma,1}
\Bigl(r - r'\Bigl) \, .
\end{eqnarray}

As for the first-order case, the use of either the function
$Q_{\alpha,\gamma}^{(2)}(q)$ defined by Eqs. (B44)-(B49)
or of the full function $X_{\alpha,\gamma}^{(2)}(q)$
of Eq. (B22) for $j=2$ leads to the same results for
the physical quantities up to second order in the
density of excited pseudoparticles.

Note that the free terms of Eqs. (B47)-(B49) involve the 
first-order functions only. This implies that the unique 
solutions of these integral equations can be expressed in 
terms of the first-order functions. Therefore, following Eqs. 
$(98)$ and (B22) the second-order functions can
also be expressed in terms of the pseudoparticle
phase shifts.

\vfill
\eject
\appendix{NORMAL-ORDERED HAMILTONIAN EXPRESSION}

In order to derive the first-order and second-order Hamiltonian
terms of Eqs. $(105)$ and $(106)$, we again consider eigenvalues 
and deviations. The energy associated with the Hamiltonian
$(83)$ reads

\begin{eqnarray}
E_{SO(4)} & = & -{2tN_a\over 2\pi}
\int_{q^{(-)}_{c,0}}^{q^{(+)}_{c,0}}N_{c,0}(q)\cos K(q)
+ {4tN_a\over 2\pi}\sum_{\gamma =1}\int_{q^{(-)}_{c,
\gamma}}^{q^{(+)}_{c,\gamma}}N_{c,\gamma}(q)
Re \sqrt{1-[u(R_{c,\gamma}(q)-i\gamma)]^2}
\nonumber \\
& + & U[{N_a\over 4} - {N_{c,0}\over 2}
- \sum_{\gamma =1}\gamma N_{c,\gamma}] \, .
\end{eqnarray}
To caculate the bands $(107)-(109)$ we introduce in 
this energy Eqs. (B13), (B14), and (B21) and expand the obtained 
expression to first order in the deviations with the result

\begin{eqnarray}
\Delta E^{(1)}_{SO(4)} & = & {N_a\over 2\pi}\{
\int_{q^{(-)}_{c,0}}^{q^{(+)}_{c,0}} \delta N_{c,0}(q)[
- 2t\cos K^{(0)}(q) - {U\over 2}]
\nonumber \\
& + & \sum_{\gamma =1}\int_{q^{(-)}_{c,\gamma}}^{q^{(+)}_{c,\gamma}}
\delta N_{c,\gamma}(q)[4t
Re \sqrt{1-[u(R^{(0)}_{c,\gamma}(q)-i\gamma)]^2} -\gamma U]
\nonumber \\
& + & 2t\int_{Q^{(-)}}^{Q^{(+)}}dk
\widetilde{Q}^{(1)}(k)\sin k
- 4t\sum_{\gamma =1}\theta (N_{c,\gamma})[\int_{-\infty}^{r^{(-)}_{c,\gamma}}
+ \int_{r^{(+)}_{c,\gamma}}^{\infty}]dr
\bar{Q}^{(1)}_{c,\gamma}(r)
Re\Bigl({u^2[r - i\gamma]\over \sqrt{1 - u^2[r - i\gamma]^2}}\Bigl)\} 
\, ,
\end{eqnarray}
where the functions $\widetilde{Q}^{(1)}(k)$ and
$\bar{Q}^{(1)}_{c,\gamma}(r)$ are defined by Eqs (B26)-(B28).
As in the case of the second term of the rhs of Eq. (B22),
up to second scattering order (and to second order in the 
density of excited pseudoparticles) the last term of Eq. (C2) does 
not contribute to the energy expression. Therefore, it can
be omitted. The use of Eqs. (B22), (B24), (B26), and $(98)$ 
in (C2) then leads after some straightforward algebra to

\begin{equation}
\Delta E^{(1)}_{SO(4)}= \sum_{q,\alpha,\gamma}
\epsilon^0_{\alpha ,\gamma}(q)\delta \hat{N}_{\alpha ,\gamma}(q)
\, ,
\end{equation}
with the bands given by Eqs. $(107)-(109)$ in terms of the
phase shifts (B30)-(B40).

To obtain the equivalent band expressions $(110)-(112)$ requires 
introducing in the energy expression (C2) the functions defined by 
Eqs. (B26)-(B29) and performing some integrations by using symmetry 
properties of the obtained integral-equation kernels. After some 
algebra we find again the energy expression (C3) with the 
pseudoparticle bands given by

\begin{equation}
\epsilon_{c,0}^0(q) = -{U\over 2} -2t\cos K^{(0)}(q) 
- \int_{r^{(-)}_{s,0}}^{r^{(+)}_{s,0}}dr 2t\eta_{s,0}(r)
{1\over \pi}\tan^{-1}\left(r -R^{(0)}_{c,0}(q)\right) \, ,
\end{equation}

\begin{equation}
\epsilon_{c,\gamma}^0(q) = -\gamma U 
+ 4t Re \sqrt{1 - u^2[R^{(0)}_{c,\gamma}(q) - i\gamma]^2}
- \int_{Q^{(-)}}^{Q^{(+)}}dk{1\over \pi}
\tan^{-1}\left({{\sin k\over u} -R^{(0)}_{c,\gamma}(q)
\over\gamma}\right)2t\eta_{c,0}(k) \, ,
\end{equation}
and

\begin{eqnarray}
\epsilon_{s,\gamma}^0(q) & = & 
- 2t\int_{Q^{(-)}}^{Q^{(+)}}dk{1\over \pi}
\tan^{-1}\left({{\sin k\over u} -R^{(0)}_{s,\gamma}(q)
\over 1+\gamma}\right)\sin k \nonumber \\
& + & \int_{r^{(-)}_{s,0}}^{r^{(+)}_{s,0}}dr 2t\eta_{s,0}(r)
\left[{\Theta_{1,1+\gamma}\Bigl(r-R^{(0)}_{s,\gamma}(q)\Bigl)
\over 2\pi} - \int_{r^{(-)}_{c,0}}^{r^{(+)}_{c,0}}dr'
{\tan^{-1}\left({r' -R^{(0)}_{s,\gamma}(q)\over 1+\gamma}\right)
\over \pi^2[1 + (r- r')^2]}\right] \, ,
\end{eqnarray}
where the functions $2t\eta_{c,0}(k)$ and $2t\eta_{\alpha,\gamma}(r)$
are defined by the integral equations  

\begin{equation}
2t\eta_{c,0}(k) = 2t\sin k 
+ {\cos k\over u} \int_{r^{(-)}_{s,0}}^{r^{(+)}_{s,0}}dr 
{2t\eta_{s,0}(r)\over \pi\left[1+(r - {\sin k\over u})^2\right]} \, ,
\end{equation}

\begin{equation}
2t\eta_{c,\gamma}(r) =  
- 4t Re\Bigl({u^2[r - i\gamma]\over
\sqrt{1 - u^2[r - i\gamma]^2}}\Bigl)
+ \int_{Q^{(-)}}^{Q^{(+)}}dk {2t\eta_{c,0}(k)\over
\pi\gamma\left[1+({{\sin k\over u} -r\over\gamma})^2\right]} \, ,
\end{equation}
and

\begin{equation}
2t\eta_{s,\gamma}(r) = 
\int_{Q^{(-)}}^{Q^{(+)}}dk{2t\eta_{c,0}(k)\over 
\pi(1+\gamma)\left[1+({{\sin k\over u} - r\over 1+\gamma})^2\right]} 
- \int_{r^{(-)}_{s,0}}^{r^{(+)}_{s,0}}dr' 2t\eta_{s,0}(r')
{\Theta^{[1]}_{1,1+\gamma}\Bigl(r'- r\Bigl)
\over 2\pi} \, .
\end{equation}

The use of these equations and comparision of the band expressions 
(C4)-(C6) with the derivatives of the functions (C7)-(C9) readly 
leads to the simple expressions $(110)-(112)$.

In order to derive the expression for the second-order Hamiltonian 
$(106)$ and associate $f$ functions $(117)$ we expand the energy 
(C1) to second-order with the result 

\begin{eqnarray}
\Delta E^{(2)}_{SO(4)} & = & {N_a\over {2\pi}}\{
\int_{q_{c,0}^{(-)}}^{q_{c,0}^{(+)}} dq \delta N_{c,0}(q)
{2t\sin K^{(0)}(q)\over 2\pi\rho_{c,0}\Bigl(K^{(0)}(q)\Bigl)}  
Q_{c,0}^{(1)}(q)
\nonumber\\
& - & \sum_{\gamma =1}\int_{q^{(-)}_{c,\gamma}}^{q^{(+)}_{c,\gamma}}
dq \delta N_{c,\gamma}(q){Q^{(1)}_{c,\gamma}(q)\over
2\pi\rho_{c,\gamma}\Bigl(R^{(0)}_{c,\gamma}(q)\Bigl)}
4t Re\Bigl({u^2[R^{(0)}_{c,\gamma}(q) - i\gamma]
\over \sqrt{1 - u^2[R^{(0)}_{c,\gamma}(q) - i\gamma]^2}}\Bigl) 
\nonumber\\
& + & {2t\sin Q\over 2\pi\rho_{c,0}(Q)}{1\over 2}
\sum_{j=\pm 1}[Q_{c,0}^{(1)}(jq_{Fc,0})]^2
\nonumber\\
& + & \sum_{\gamma =1}{\theta (N_{c,\gamma})\over 2}\sum_{j=\pm 1}
{[Q^{(1)}_{c,\gamma}(jq_{c,\gamma})]^2\over
2\pi\rho_{c,\gamma}(r_{c,\gamma})}
4t Re\Bigl({u^2[r_{c,\gamma} - i\gamma]
\over \sqrt{1 - u^2[r_{c,\gamma} - i\gamma]^2}}\Bigl) 
\nonumber\\
& + & \int_{Q^{(-)}}^{Q^{(+)}}dk\widetilde{Q}^{(2,*)}(k)
2t\sin k
\nonumber\\
& - & \sum_{\gamma =1}\theta (N_{c,\gamma})[\int_{-\infty}^{r^{(-)}_{c,\gamma}}
+ \int_{r^{(+)}_{c,\gamma}}^{\infty}]
dr\bar{Q}^{(2,*)}_{c,\gamma}(r)
4t Re\Bigl({u^2[r - i\gamma]
\over \sqrt{1 - u^2[r - i\gamma]^2}}\Bigl)\} 
\, .
\end{eqnarray}
Again, the last term of the rhs of this equation does
not contribute to the energy expression to second scattering
order (and to second order in the density of excited
pseudoparticles).
Inserting the suitable functions in the rhs of Eq. (C14), performing 
some integrations by using symmetry properties of the kernels of
the integral equation (B27)-(B29) and (B47)-(B49), and replacing 
deviations by pseudomomentum normal-ordered operators $(84)$ we 
find after some algebra expression $(116)$. Note that replacing in 
Eq. $(116)$ the function $X^{(1)}_{\alpha,\gamma}(q)$ by the associate
function $Q^{(1)}_{\alpha,\gamma}(q)$ leads to the same result. 
By the use of Eq. $(98)$ expression $(116)$ can be rewritten in terms 
of the $f$ functions $(117)$ as given in the rhs of Eq. $(106)$.



\begin{references}
\bibitem[1]{Solyom}
        J. S\'olyom, Adv. Phys. {\bf 28}, 201 (1979);
        F. D. M. Haldane, J. Phys. C {\bf 14},
        2585 (1981).       
\bibitem[2]{Meden}
        V. Meden and K. Sch\"onhammer, Phys. Rev. B {\bf 46},
        15 753 (1992); J. Voit {\it ibid.} {\bf 47}, 6740 (1993);             
        Peter Kopietz, Volker Meden, and Kurt Sch\"onhammer, 
        Phys. Rev. Lett. {\bf 74}, 2997 (1995). 
\bibitem[3]{Pines}
        D. Pines and P. Nozi\`eres, in {\em The Theory of 
        Quantum Liquids},
        (Addison-Wesley, Redwood City, 1966 and 1989), Vol. I.
\bibitem[4]{Baym}
        Gordon Baym and Christopher J. Pethick, in
        {\em Landau Fermi-Liquid Theory Concepts and Applications},
        (John Wiley \& Sons, New York, 1991).
\bibitem[5]{Anderson}      
        Philip W. Anderson, Phys. Rev. Lett. {\bf 64},
        1839 (1990); {\bf 65} 2306 (1990);
        P. W. Anderson and Y. Ren, in {\it High Temperature
        Superconductivity}, edited by K. S. Bedell,
        D. E. Meltzer, D. Pines, and J. R. Schrieffer
        (Addison-Wesley, Reading, MA, 1990).
\bibitem[6]{Essler}
        Fabian H. L. Essler and Vladimir E. Korepin, 
        Phys. Rev. Lett. {\bf 72}, 908 (1994);
        {\it ibid.} Nucl. Phys. B {\bf 426}, 505 (1994).             
\bibitem[7]{Carmelo92}
        J. M. P. Carmelo and P. Horsch,
        Phys. Rev. Lett. {\bf 68}, 871 (1992).
\bibitem[8]{Carmelo93}
        J. M. P. Carmelo and A. H. Castro Neto,
        Phys. Rev. Lett. {\bf 70}, 1904 (1993).
\bibitem[9]{Carmelo94}
        J. M. P. Carmelo, A. H. Castro Neto, and 
        D. K. Campbell, Phys. Rev. Lett. {\bf 73}, 926
        (1994); {\it ibid} {\bf 74} (E), 3089 (1995); 
        J. M. P. Carmelo and A. H. Castro Neto, Phys. Rev. B
        {\bf 54}, 9960 (1996).
\bibitem[10]{Belavin}
        A. A. Belavin, A. M. Polyakov, and A. B.
        Zamolodchikov, J. Stat. Phys. {\bf 34},
        763 (1984); Nucl. Phys. B {\bf 241}, 333 (1984).             
\bibitem[11]{Blote}
        H. W. Bl\"ote, John L. Cardy, and M. P.
        Nightingale, Phys. Rev. Lett. {\bf 56}, 742
        (1985).
\bibitem[12]{Affleck}
        Ian Affleck, Phys. Rev. Lett. {\bf 56}, 746  
        (1985).
\bibitem[13]{Frahm}
        Holger Frahm and V. E. Korepin, Phys. Rev. B {\bf 42},
        10 553 (1990); {\it ibid.} {\bf 43}, 5653 (1991).             
\bibitem[14]{Bethe}
        H. A. Bethe, Z. Phys. {\bf 71}, 205 (1931).
\bibitem[15]{Yang}
        For one of the first generalizations of the Bethe
        ansatz to multicomponent systems see
        C. N. Yang, Phys. Rev. Lett. {\bf 19}, 1312
        (1967).
\bibitem[16]{Thacker}
        For a classical review of the Bethe ansatz and integrability
        in field theory and statistical mechanics see H. B. Thacker,
        Rev. Mod. Phys. {\bf 53}, 253 (1981).
\bibitem[17]{Korepinrev}
        V. E. Korepin, N. M. Bogoliubov, and A. G. Izergin,
        {\it Quantum Inverse Scattering Method and Correlation Functions}
        (Cambridge University Press, 1993).
\bibitem[18]{Lieb}
        Elliott H. Lieb and F. Y. Wu, Phys. Rev. Lett. {\bf 20},
        1445 (1968).       
\bibitem[19]{Takahashi}
        M. Takahashi, Prog. Theor. Phys. {\bf 47}, 69 (1972).
\bibitem[20]{Carmelo96}
        J. M. P. Carmelo and N. M. R. Peres, Nucl. Phys. B
        {\bf 458}, 579 (1996).
\bibitem[21]{Heilmann}
        O. J. Heilmann and E. H. Lieb, Ann. N. Y.
        Acad. Sci. {\bf 172}, 583 (1971);
        E. H. Lieb, Phys. Rev. Lett. {\bf 62},
        1201 (1989).       
\bibitem[22]{Yang89}
        C. N. Yang, Phys. Rev. Lett. {\bf 63}, 2144
        (1989); C. N. Yang and S. C. Zhang,
        Mod. Phys. Lett. B {\bf 4}, 759 (1990).
\bibitem[23]{Korepin}
        Fabian H. L. Essler, Vladimir E. Korepin, and
        Kareljan Schoutens, Phys. Rev. Lett. {\bf 67},
        3848 (1991); Nucl. Phys. B {\bf 372}, 559 (1992).             
\bibitem[24]{Ostlund}
        Stellan \"Ostlund, Phys. Rev. Lett. {\bf 69},
        1695 (1992).                                          
\bibitem[25]{Carmelo96b}
        J. M. P. Carmelo, N. M. R. Peres, and D. K.
        Campbell, unpublished;
        N. M. R. Peres, J. M. P. Carmelo, D. K. Campbell,
        and A. W. Sandvik, Z. Phys. B {\bf 103}, 217 (1997).
\bibitem[26]{Carmelo96d}
        J. M. P. Carmelo, preprint.
\bibitem[27]{Faddeev}
        L. D. Faddeev and L. A. Takhtajan, 
        Phys. Lett. {\bf 85A}, 375 (1981).             
\bibitem[28]{Carmelo95}
        J. M. P. Carmelo and N. M. R. Peres, Phys. Rev. B
        {\bf 51}, 7481 (1995).
\bibitem[29]{Haldane91}
        F. D. M. Haldane, Phys. Rev. Lett. {\bf 67},
        937 (1991).       
\bibitem[30]{Carmelo96c}
        J. M. P. Carmelo and A. H. Castro Neto, Phys. Rev. B
        {\bf 54}, 11 230 (1996).
\bibitem[31]{Carmelo90}
        J. Carmelo and A. A. Ovchinnikov, Carg\`ese Lecture
        1990 (unpublished); J. Phys.: Condens. 
        Matter {\bf 3}, 757 (1991).
\bibitem[32]{Carmelo91}
        J. Carmelo, P. Horsch, P.-A. Bares, and A. A. Ovchinnikov, 
        Phys. Rev. B {\bf 44}, 9967 (1991).
\bibitem[33]{Carmelo92b}
        J. M. P. Carmelo, P. Horsch, and A. A. Ovchinnikov, 
        Phys. Rev. B {\bf 45}, 7899 (1992).    
\bibitem[34]{Carmelo92c}
        J. M. P. Carmelo, P. Horsch, and A. A. Ovchinnikov, 
        Phys. Rev. B {\bf 46}, 14 728 (1992).    
\bibitem[35]{Carmelo93b}        
        J. M. P. Carmelo, P. Horsch, D. K. Campbell, and
        A. H. Castro Neto, Phys. Rev. B {\bf 48}, 4200 (1993);
        J. M. P. Carmelo, F. Guinea, and P. D. Sacramento, Phys. 
        Rev. B {\bf 55}, 7565 (1997).
\bibitem[36]{Carmelo94b}
        J. M. P. Carmelo, A. H. Castro Neto, and 
        D. K. Campbell, Phys. Rev. B {\bf 50},
        3667 (1994); 3683 (1994).
\bibitem[37]{Anto}
        A. H. Castro Neto, H. Q. Lin, Y. -H.
        Chen, and J. M. P. Carmelo, Phys. Rev. B {\bf 50}
        14 032 (1994).
\bibitem[38]{Korepin79}
        V. E. Korepin, Teor. Mat. Fiz. {\bf 41}, 169
        (1979) [Theor. Math. Phys. {\bf 41}, 953 (1980)].
\bibitem[39]{Zamo}
        Alexander B. Zamolodchikov and Alexey B. Zamolodchikov,
        Nucl. Phys. B {\bf 133}, 525 (1978).          
\bibitem[40]{Andrei}
        N. Andrei and J. H. Lowenstein, Phys. Lett.
        {\bf 91B}, 401 (1980).
\end{references}
\end{document}